\newif\ifconfver
\confverfalse      

\newif\ifcutshort      
\cutshorttrue

\newif\ifcutshortlvltwo  
\cutshortlvltwofalse

\ifconfver
    \documentclass[10pt,twocolumn,twoside]{IEEEtran}
\else
    \documentclass{IEEEtran}
\fi

\usepackage{cite}
\usepackage{graphicx}
\usepackage{psfrag}
\usepackage{url}
\usepackage{stfloats}
\usepackage{amsfonts,amssymb,amsmath,bm,paralist,theorem,cite,ifthen}
\usepackage{array}
\usepackage{caption}
\usepackage{cases}
\usepackage{calc}

\usepackage{longtable}
\usepackage{graphics,booktabs,color,epsfig,enumerate}
\usepackage{multirow}
\usepackage{algorithm,algpseudocode}
\usepackage{caption}
\usepackage{subfig} 

\usepackage{epstopdf}
\usepackage{array}
\usepackage{stfloats}
\usepackage{mathrsfs}

\usepackage{enumitem}

\newtheorem{proposition}{Proposition}

\theorembodyfont{\rmfamily}

\newtheorem{rmk}{Remark}

{\begin{list}{}{
    \settowidth{\labelwidth}{\mbox{\textnormal{#1}}}%
    \setlength{\leftmargin}{\labelwidth+\labelsep}}}%
{\end{list}}

\makeatletter
\def\changeBibColor#1{%
  \in@{#1}{}
  \ifin@\color{blue}\else\normalcolor\fis
}

\begin{document}

\bibliographystyle{IEEEtran}

\newcommand\bcc[2][c]{\ensuremath{\bm{\mathcal{#2}}}}      
\newcommand\bcl[2][c]{\ensuremath{\bm{#2}}}
\newcommand\Ib{\ensuremath{{\bm I}}}
\newcommand\Vb{\ensuremath{{\bm V}}}
\newcommand\vb{\ensuremath{{\bm v}}}
\newcommand\Hb{\ensuremath{{\bm H}}}
\newcommand\ub{\ensuremath{{\bm u}}}
\newcommand\hb{\ensuremath{{\bm h}}}
\newcommand\xb{\ensuremath{{\bm x}}}
\newcommand\gb{\ensuremath{{\bm g}}}
\newcommand\Thetab{\ensuremath{{\bm \Theta}}}
\newcommand\thetab{\ensuremath{{\boldsymbol \theta}}}
\newcommand\Gb{\ensuremath{{\bm G}}}
\newcommand\Pb{\ensuremath{{\bm P}}}
\newcommand\taub{\ensuremath{{\boldsymbol \tau}}}
\newcommand\zb{\ensuremath{{\bm z}}}
\newcommand\st{\ensuremath{{\rm ~s.t.}}}


\def\blue{\color{blue}}
\def\red{\color{red}}
\definecolor{orange}{RGB}{255,107,0}
\def\orange{\color{orange}}

\title{Symbol-Level Precoding for Integrated Sensing and Covert Communication}

\ifconfver \else {\linespread{1.1} \rm \fi

\author{ Yufei Wang, Qiang Li, Hongli Liu, Ying Zhang and Jingran Lin
	\thanks{Y. Wang, Q. Li, H. Liu, Y. Zhang and J. Lin are with
		School of Information and Communication  Engineering, University of Electronic Science and Technology of China, Chengdu, P.~R.~China, 611731. Q. Li and J. Lin are also affiliated with the Tianfu Jiangxi Laboratory, Chengdu 641419, China.
}
\thanks{This work was supported  by the National Natural Science Foundation of China
	under Grant 62171110.}
	\thanks{Corresponding author: Qiang Li,  E-mail: lq@uestc.edu.cn.}
}
\maketitle

\begin{abstract}
	Integrated sensing and communication (ISAC) systems have emerged as a promising solution to improve spectrum efficiency and enable functional convergence. However, ensuring secure information transmission while maintaining high-quality sensing performance remains a significant challenge. In this paper, we investigate an integrated sensing and covert communication (ISCC) system, in which a base station (BS) simultaneously serves multiple downlink users and senses malicious targets that may act as both potential eavesdroppers (Eves) and wardens. We propose a novel symbol-level precoding (SLP)-based waveform design for ISCC that achieves covert communication intrinsically, without requiring additional transmission resources such as artificial noise. The proposed design integrates  symbol shaping to enhance reliability for legitimate users and noise shaping to obscure transmission activities from the targets.  For imperfect channel state information (CSI), the framework incorporates bounded uncertainty models for user channels and target angles, yielding a more robust design. The resulting ISCC waveform optimization problem is non-convex; to address this, we develop a low-complexity proximal distance algorithm (PDA) with closed-form updates under both PSK and QAM modulations. 
	Simulation results demonstrate that the proposed method achieves superior covertness and sensing-communication performance with negligible degradation compared to traditional beamforming and conventional SLP approaches without noise-shaping mechanisms.
	
\end{abstract}


\ifconfver \else \IEEEpeerreviewmaketitle} \fi

\section{Introduction}

Integrated sensing and communication (ISAC) has recently emerged as a key technology to address the growing demand for spectrum efficiency and functional convergence in future wireless systems. ISAC unifies communication and sensing functionalities within a common wireless platform by sharing frequency bands and hardware resources, yielding significant benefits such as improved spectral efficiency, reduced system cost, and mutual performance enhancement~\cite{liu2022integrated}. In particular, combining multiple-input multiple-output (MIMO) techniques with ISAC has attracted significant interest, as the spatial degrees of freedom offered by MIMO arrays enable joint optimization of communication and sensing capabilities~\cite{liu2018toward}.

Despite its advantages, ISAC introduces fundamental challenges, especially in waveform design~\cite{lu2024integrated}, due to the need to simultaneously meet distinct performance metrics—such as data rate, bit error rate, range resolution, and detection probability~\cite{liu2022integrated,liu2018toward}. Unlike traditional systems where communication and sensing are developed separately, ISAC systems must balance these conflicting requirements within a unified design framework under shared spectral and hardware constraints~\cite{meng2024cooperative}. Waveform design for MIMO-ISAC systems has thus become a vibrant area of research, with approaches ranging from sensing-centric designs that embed communication data into radar waveforms~\cite{hassanien2016phase,chen2011novel}, to communication-centric designs that repurpose communication waveforms like OFDM for sensing purposes~\cite{sturm2009joint,shi2022device}, and integrated designs that jointly optimize both functions from the ground up~\cite{zheng2019radar,liu2020joint,luong2021radio,zhang2021enabling,zhao2024joint}. Among the integrated strategies, symbol-level precoding (SLP) has emerged as a promising solution due to its ability to finely control transmitted signals, thereby improving both communication reliability and sensing precision.

SLP exploits constructive interference (CI) by leveraging symbol knowledge and multiuser interference (MUI) to push received signals away from constellation decision boundaries, enhancing detection reliability~\cite{masouros2009dynamic,masouros2015exploiting}. Unlike traditional precoding methods that statistically suppress interference, SLP converts MUI into a constructive component, thereby improving quality of service (QoS) in multiuser systems~\cite{alodeh2018symbol,li2020tutorial}. Additionally, symbol-level control allows finer beampattern shaping, which can further enhance sensing performance. These advantages have recently motivated efforts to extend SLP to ISAC systems for simultaneous enhancement of communication and sensing~\cite{liu2021dual}.

Alongside performance enhancement, ensuring secure information transmission has become increasingly critical in ISAC systems~\cite{deligiannis2018secrecy}. The dual-purpose nature of ISAC waveforms introduces unique security challenges, as sensing targets may also act as potential eavesdroppers (Eves). Physical layer security (PLS), rooted in information-theoretic principles~\cite{wyner1975wire}, has been proposed as a promising approach by exploiting the spatial degrees of freedom in multi-antenna systems~\cite{bloch2011physical,deligiannis2018secrecy,jiang2023physical,cao2024sensing,su2025secure}. Techniques such as secure beamforming and artificial noise (AN) injection have been explored to enhance secrecy. For instance, \cite{su2020secure} investigates secure ISAC systems where radar targets act as eavesdroppers, employing AN to interfere with their reception and improve secrecy rate.

However, existing PLS techniques face practical limitations. First, information-theoretic security requires long random codes, which are difficult to implement. Second, PLS does not conceal the existence of communication—eavesdroppers may still detect ongoing transmissions even if they cannot decode them. Compared with PLS, covert communication has emerged as a higher-level paradigm, aiming not only to secure the transmitted information but also to hide the very presence of transmission~\cite{yan2019low}, motivating the study of covert communication techniques in ISAC systems~\cite{forouzesh2020covert,hu2023covert,tang2024dual,qian2025two,luo2023ris,zhu2024active,guo2025joint}.  
Injecting jamming signals has been shown to be a powerful way to enhance covertness. In~\cite{forouzesh2020covert}, a jammer was employed to improve the covert transmission rate by performing null-space beamforming with the intended receiver placed in the null space. The authors in~\cite{hu2023covert} further analyzed the trade-off between covert rate and communication performance, demonstrating that AN injection is critical to achieve covertness but inevitably induces a rate loss. More recently, a two-stage ISAC framework was proposed in~\cite{qian2025two}, where Bob first performs ISAC-based detection to identify the presence of the warden; based on this outcome, Alice adapts her covert transmission, while Bob dynamically decides whether to transmit AN.  
Reconfigurable intelligent surfaces (RIS) have also been investigated as an effective enabler for covert ISAC by reconfiguring the wireless propagation environment to improve covertness~\cite{luo2023ris}. Active RIS can optimize reflection coefficients to redistribute energy, enhancing signals received by legitimate users and sensing targets while suppressing those at the warden~\cite{zhu2024active}. More advanced architectures, such as active STAR-RIS, support simultaneous reflection and transmission with amplification, enabling AN-like interference that confuses the warden and further improves covert performance~\cite{guo2025joint}. 

These studies reveal that while both AN- and RIS-assisted covert ISAC schemes are effective, they follow the traditional principle of embedding information signals within external interference or background noise to degrade the warden’s detection capability by inducing missed detections or false alarms~\cite{ma2022covert}. However, such approaches typically require additional hardware support (e.g., dedicated jammers or reconfigurable surfaces), which increases system complexity and deployment cost.
In addition, key-based physical-layer encryption (PLE) techniques, such as elliptic curve modulation~\cite{oikonomou2025ecm} or dynamic constellation rotation~\cite{hou2022ple}, obfuscate the constellation mapping through secret keys and thus provide another line of defense. Yet, they still depend on key generation and sharing mechanisms.

In this paper, we investigate an  integrated sensing and covert communication (ISCC) system where a base station (BS) simultaneously serves multiple downlink users and detects malicious targets that threaten the communication links as both Eves and wardens. We propose a novel symbol-level ISCC waveform design that achieves covert communication {\it intrinsically}, without the need for AN or other additional transmission resources. 	
Leveraging the spatial degrees of freedom offered by MIMO systems, the proposed design employs  ``symbol shaping'' --- defined as the per-symbol optimization of a complex-valued transmit signal vector --- to realize ISCC. Specifically,  this design takes as input the desired symbol of each communication user (from a fixed modulation constellation such as PSK or QAM) and the instantaneous channel realization, and determines the transmit vector to fulfill two key shaping objectives:
\begin{itemize}
	\item {\bf Symbol shaping:} Steers the received signal at each intended communication user away from modulation decision boundaries and aligns it precisely with the desired constellation point, thereby enhancing symbol detection reliability.
	\item {\bf Noise shaping:} Shapes the transmitted waveform to be statistically indistinguishable from Gaussian noise at unintended malicious targets, thereby concealing communication activities and achieving covertness.
\end{itemize}

To this end, we formulate an optimization problem that maximizes the worst-case signal-to-clutter-plus-noise ratio (SCNR) at the targets, while enforcing symbol error probability (SEP) constraints for communication users. A noise-shaping constraint is introduced to ensure covertness. Due to the non-convexity of the problem, we adopt a sampling-based reformulation and develop a low-complexity proximal distance algorithm (PDA) to solve it efficiently. The main contributions of this paper are summarized as follows:
\begin{enumerate}
	\item We introduce a novel SLP-based ISCC framework that achieves intrinsic covertness by directly shaping transmit vectors—realizing reliable communication at users while making the received waveform at malicious targets resemble Gaussian noise. Unlike conventional covert communication, our approach requires no additional artificial noise or interference, offering an endogenous covertness mechanism. 
	
	\item We formulate the ISCC transmit optimization problems for both PSK and QAM modulations, which jointly enforce sensing, SEP, covertness, and power constraints. To address their non-convexity, we develop a customized MM–PDA algorithm with closed-form updates, yielding significantly lower complexity than generic solvers.  
	
	\item We extend the framework to imperfect CSI by incorporating bounded uncertainty sets for user channels and target angles. This robust ISCC design guarantees worst-case performance, and simulations confirm that it preserves covertness and reliability while outperforming beamforming and SLP baselines.  
\end{enumerate}

\textit{Organization:} Section~\ref{sec:system_model} introduces the system model and formulates the symbol-level ISCC problem. Section~\ref{sec:method} develops an iterative PDA  method for the problem under the PSK modulation. Section~\ref{sec:QAM} generalizes the formulation to general QAM modulation and adapts PDA to QAM-specific constraints. Section~\ref{sec:robust} extends the proposed framework to robust ISCC design under imperfect CSI. Section \ref{sec:num_results} presents numerical results to evaluate the performance of the proposed approach. Finally, Section \ref{sec:conclusions} concludes the paper.

\textit{Notations:} Throughout this paper, boldface lowercase letters (e.g., $\bm{x}$) denote vectors, and boldface uppercase letters (e.g., $\bm{H}$) denote matrices. The superscripts $(\cdot)^T$ and $(\cdot)^H$ represent the transpose and Hermitian transpose, respectively. The notation $\Re(\cdot)$ and $\Im(\cdot)$ denote the real and imaginary parts of a complex number, respectively. The operator $\otimes$ denotes the Kronecker product. For a complex Gaussian vector $\bm{x} \in \mathbb{C}^n$ with mean $\bm{\mu}$ and covariance matrix $\bm{\Sigma}$, we write $\bm{x} \sim \mathcal{CN}(\bm{\mu}, \bm{\Sigma})$. The Q-function is defined as $Q(x) = \frac{1}{\sqrt{2\pi}} \int_x^\infty e^{-t^2/2} dt$, and $Q^{-1}(\cdot)$ denotes its inverse. We use $\|\cdot\|$ to denote the Euclidean norm and $\|\cdot\|_F$ for the Frobenius norm. In addition, the key notations frequently used in this paper are listed in Table~\ref{tab:notation} for quick reference.

\begin{table}[t]
	\centering
	\caption{List of Notations}
	\begin{tabular}{p{0.18\columnwidth}p{0.75\columnwidth}}
		\toprule
		Notation & Definition \\
		\midrule
		$K_C$ & Number of clutter scatterers in the sensing environment \\
		$K_T$ & Number of sensing targets \\		
		${\cal K}_T$ & Set of sensing targets \\
		$K_U$ & Number of communication users \\
		${\cal K}_U$ & Set of communication users \\
		$N_t$ & Number of transmit antennas at the ISCC transmitter \\
		$N_r$ & Number of receive antennas at the ISCC receiver \\
		$L$ & Number of symbols per CPI \\				
		$M$ &  PSK/QAM modulation order\\
		$T_p$ & Pulse duration \\
		$\mathbf{h}_k$ & Channel vector from the transmitter to user $k$ \\
		$\mathbf{x}_\ell$ & Transmit signal vector in the $\ell$-th symbol interval \\
		$s_{k,\ell}$ & PSK/QAM symbol for user $k$ in the $\ell$-th symbol interval \\
		$y_{k,\ell}$ & Received signal at user $k$ in the $\ell$-th symbol interval \\
		$y_{t,\ell}$ & Received signal at target $t$ in the $\ell$-th symbol interval \\
		$P$ & Maximum transmit power budget \\
		$\bm{A}_k$ & Spatio-temporal steering matrix for target $k$ \\
		$\bm{A}_c$ & Spatio-temporal steering matrix for clutter scatterer $c$ \\	
        $\bm a(\theta)$ & Array steering vector corresponding to angle $\theta$ \\
		\bottomrule
	\end{tabular}
	\label{tab:notation}
\end{table}

\section{System Model and Problem Formulation} \label{sec:system_model}

\subsection{Signal Model}
\begin{figure}[!h]
\centerline{\resizebox{.4\textwidth}{!}{\includegraphics{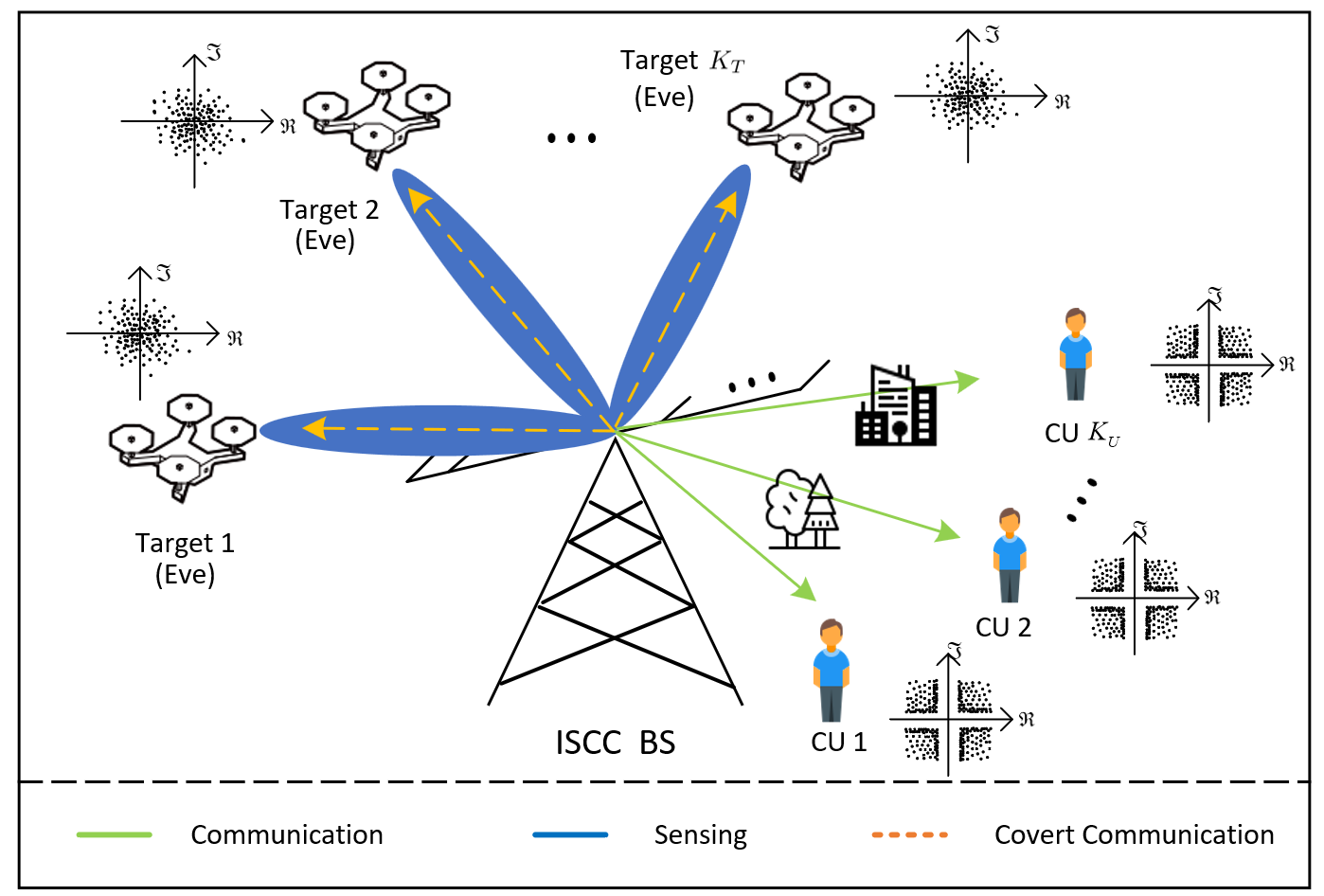}}
}   \caption{An integrated sensing and covert communication system.} \label{fig:system_model}
\end{figure}

As shown in Fig.~\ref{fig:system_model}, an ISAC transmitter with $N_t$ transmit and $N_r$ receive antennas (both arranged as uniform linear arrays (ULAs) with spacing $d$ and wavelength $\lambda$) serves $K_U$ single-antenna communication users (CUs) while detecting $K_T$ malicious targets.  Each malicious target is also treated as a potential Eve (i.e., the warden) that may attempt to intercept CUs' transmissions. To ensure security, we adopt a covert communication framework and design ISCC waveforms that simultaneously enable reliable communication, effective sensing, and concealment of communication activity from the malicious targets.

In the following, we will describe the  performance metrics for sensing and communications and formulate the ISCC problem.

\subsection{Radar Sensing Performance Metric}

Let $w_i(t)$ be the transmit signal at the $i$-th antenna 
\begin{equation}
    w_i(t) = \sum_{\ell=1}^L x_{i,\ell} p(t-(\ell-1)T_p), \quad i=1,\ldots, N_t, 
\end{equation}
where $p(t)$ is the signature  of each transmit pulse, $T_p$ is the pulse repetition interval (PRI), $L$ is the number of pulses within a 
coherent processing interval (CPI), and $ x_{i,\ell} \in \mathbb{C}$ is  the codedword modulating the $\ell$-th pulse at the $i$-th antenna.

The received signal \( y_j(t) \) at the \( j \)-th receive antenna of the ISAC system, including both target and clutter returns, can be expressed as 
\begin{equation}
    \begin{aligned}
y_j(t) = \sum_{i=1}^{N_t} \sum_{\ell=1}^L &   x_{i,\ell} \left[ \sum_{k=1}^{K_T} \alpha_{j,i,k} p(t - \tau_k - (\ell-1)T_p) \right.  \\
& \hspace{-1cm} \left. + \sum_{c=1}^{K_C} \beta_{j,i,c} p(t - \tau_c - (\ell-1)T_p) \right] + n_j(t),
\end{aligned}
\end{equation}
where  $K_C$ denotes the number of clutters,  the delays \( \tau_k \) and \( \tau_c \) correspond to the propagation time for the target and clutter echoes, respectively,  \( n_j(t) \sim \mathcal{CN}(0, \sigma_0^2) \) is additive white Gaussian noise at the receiver.
The \( \alpha_{j,i,k} \in \mathbb{C} \) represents the backscattering and propagation effects associated with the \( k \)-th target, while \( \beta_{j,i,c} \) is the clutter gain for the \( c \)-th scatterer. When specifying to the ULA channel with  the azimuth direction of the target at $\theta_k$, the  $\alpha_{j,i,k}$ can be expressed as 
\begin{equation}
    \alpha_{j,i,k} = [ \nu_k \bm a_r(  \theta_k) \bm a_t^H( \theta_k)]_{j,i},\quad \beta_{j,i,c} = [ \nu_c \bm a_r(  \theta_c) \bm a_t^H( \theta_c)]_{j,i},
\end{equation}
where $\nu_k$ with $\mathbb{E}\{|\nu_k|^2\} = \varsigma_k^2$ and $\nu_c $ with $\mathbb{E}\{|\nu_c|^2\} = \varsigma_c^2$ account for the path loss and the Radar Cross Section (RCS) of the targets and clutters, respectively, and 
\begin{equation}\label{eq:steer_vector}
\begin{aligned}	
	\bm a_r(  \theta) &= \frac{1}{\sqrt{N_r}} \left[ 1, e^{\mathfrak{j} 2\pi d\sin(\theta)/\lambda } \ldots,  e^{\mathfrak{j} 2\pi d(N_r-1) \sin(\theta)/\lambda }   \right], \\
	\bm a_t(  \theta) &= \frac{1}{\sqrt{N_t}} \left[ 1, e^{\mathfrak{j} 2\pi d \sin(\theta) /\lambda } \ldots,  e^{\mathfrak{j} 2\pi d(N_t-1) \sin(\theta) /\lambda}   \right].
\end{aligned}
\end{equation}

After matched filtering the received signal \( y_j(t) \) with the pulse \( p(t) \), the discrete-time baseband output at the \( j \)-th receive antenna for the \( k \)-th target can be expressed as  
\begin{equation}\label{eq:rx_sig_radar}
y_{j,k,\ell} = \sum_{i=1}^{N_t}  x_{i,\ell} \alpha_{j,i,k} + \sum_{c \in \mathcal{C}(k)} \sum_{i=1}^{N_t}  x_{i,\ell} \beta_{j,i,c} + n_{j,\ell},
\end{equation}
for all $\ell \in {\mathscr{L}}\triangleq \{1,\ldots, L\}, ~k \in {\cal K}_T$, where \( \mathcal{C}(k) \) represents the set of clutter\footnote{Although termed ``clutter,'' the second summation in \eqref{eq:rx_sig_radar} accounts for all echoes falling into the $k$-th range bin—including true clutter, sidelobe leakage after matched filtering, and other targets within the $k$-th bin—which are collectively absorbed into the interference-plus-noise covariance $\bm R_k$ in the SCNR expression~\eqref{eq:snr_radar}.} scatters whose delays fall into range bin \( k \). 
Denote by  $\bm y_{k,\ell} = [y_{1,k,\ell}, \ldots, y_{N_r,k,\ell}]^T$, $\bm y_k =\! [\bm y_{k,1}^T, \ldots, \bm y_{k,L}^T]^T$, $\bm x_\ell \!= \! [x_{1,\ell}, \ldots, x_{N_t,\ell}]^T $, $\bm x\! = \![\bm x_1^T, \ldots, \bm x_L^T]^T$, $\bm n_\ell  = [n_{1,\ell}, \ldots, n_{N_r,\ell}]^T$ and $\bm n= [\bm n_1^T, \ldots, \bm n_L^T]^T$. The model~\eqref{eq:rx_sig_radar} can be compactly written as 
\begin{equation}
\bm y_k = \nu_k \bm A_k \bm x +   \sum_{c\in {\cal C}(k)}   \nu_c  \bm A_c  \bm x  +   \bm n, ~~ k\in {\cal K}_T
\end{equation}
where $\bm A_k$ is the steering matrix, defined as
\begin{equation}
\bm A_k  = \bm I_{L} \otimes [ \bm a_r( \theta_k) \bm a_t^H( \theta_k)] \in \mathbb{C}^{N_r L \times N_t L},
\end{equation}
and $\bm A_c$ is defined similarly.

Let $\bm w_k \in \mathbb{C}^{N_r L}$ be a spatial-temporal receive  beamformer  for the $k$th target. The output after receive beamforming is given by
\begin{equation}
{r}_k= \bm w_k^H \bm y_k= \nu_k \bm w_k^H \bm A_k \bm x +   \sum_{c\in {\cal C}(k)}   \nu_c  \bm w_k^H \bm A_c \bm x + \bm w_k^H \bm n.
\end{equation}
The  output SCNR  is given by
\begin{equation}\label{eq:radar_sinr}
\gamma_k(\bm x, \bm w_k) =  \frac{\bm w_k^H \bm A_k \bm x \bm x^H \bm A_k^H  \bm w_k  }{  \bm w_k^H \bm R_k \bm w_k},
\end{equation}
where $\bm R_k =  \varsigma_k^{-2} (\sigma_0^2 \bm I +  \sum_{c\in {\cal C}(k)}  \varsigma_c^2 \bm A_c  \bm x  \bm x^H \bm A^H_c )$.  The $\gamma_k(\bm x, \bm w_k) $ is a Rayleigh ratio with respect to $\bm w_k$, and it is well-known that the optimal $\bm w_k $ that maximizes $\gamma_k(\bm x, \bm w_k) $  is  $ \bm w_k^\star= \eta {\bm R_k}^{-1} \bm A_k \bm x$ for any nonzero $\eta$, and the resultant SCNR is given by
\begin{equation} \label{eq:snr_radar}
\gamma_k(\bm x) =  \bm x^H  \bm A_k^H  \bm R_k^{-1} \bm A_k \bm x,~~ k\in {\cal K}_T.
\end{equation}

Since under the Neyman--Pearson (NP) criterion, the detection performance is directly related to the received SCNR, we adopt $\gamma_k(\bm{x})$ in~\eqref{eq:snr_radar} as the sensing performance metric and optimize $\bm{x}$ accordingly. 
For clarity, we recall that under the NP framework with a matched filter in Gaussian clutter-plus-noise, the detection probability $P_D$ and the false-alarm probability $P_{\mathrm{FA}}$ are explicitly related to the SCNR. 
After matched filtering and normalization, the decision statistic under the target-present hypothesis $\mathcal{H}_1$ follows a Gaussian distribution with mean $\sqrt{\mathrm{SCNR}}$ and unit variance, while under the target-absent hypothesis $\mathcal{H}_0$ it follows a zero-mean Gaussian distribution with unit variance. 
For a fixed false-alarm probability $P_{\mathrm{FA}}$, the NP criterion sets the detection threshold $\eta = Q^{-1}(P_{\mathrm{FA}})$, where $Q(\cdot)$ is the Gaussian $Q$-function. 
The detection probability can then be expressed as
\begin{equation}
	P_D = Q\left( Q^{-1}(P_{\mathrm{FA}}) - \sqrt{\mathrm{SCNR}} \right).
\end{equation}
This expression shows that the SCNR directly controls the separation between the $\mathcal{H}_1$ and $\mathcal{H}_0$ distributions: a higher SCNR increases the separation, thereby improving $P_D$ for a fixed $P_{\mathrm{FA}}$. 
Consequently, maximizing the worst-case SCNR is equivalent to maximizing the guaranteed target detection probability under a fixed false-alarm rate.

\subsection{Communication Performance Metric}
Similar to the radar  processing, the received baseband discrete-time signal at the $k$-th CU can be written as
\begin{equation} \label{eq:slp_model}
y_{k,\ell} = \bm h_k^H \bm x_\ell +   \tilde{n}_{k,\ell}, ~~~ \forall~\ell \in {\mathscr{L}},~k \in {\cal K}_U,
\end{equation}
where $\bm h_k = [h_{1,k}, \ldots, h_{N_t, k}]^T \in \mathbb{C}^{N_t\times 1}$, and $\tilde{n}_{k,\ell} \sim {\cal CN}( 0, \tilde{\sigma}_k^2 )$. The $y_{k,\ell} $ is fed into a demodulator ${\sf dec}(\cdot): \mathbb{C} \rightarrow {\cal S}$ and mapped into a  symbol $s_{k,\ell} \in {\cal S}$. Herein, $s_{k,\ell}$ is the desired symbol for CU $k$ at time $\ell$ and ${\cal S}$ is a pre-specified constellation set, e.g., PSK or QAM. For simplicity, we will first consider the PSK constellation and  the more general QAM case will be studied in Section~\ref{sec:QAM}. Considering $M$-PSK constellation
\begin{equation}
    { \cal S} \triangleq  \{ s~|~ s=e^{\mathfrak{j} n2\pi/M}, ~n=0,\ldots, M-1\},
\end{equation}
the demodulator  is given by ${\sf dec}(y)= e^{\mathfrak{j}\hat{n}2\pi/M}$, where $\hat{n}\in \{0, \ldots, M-1\}$ satisfies $\angle y \in [2\pi \hat{n}/M- \pi/M, ~2\pi \hat{n}/M +\pi/M]$.

Following the principle of SLP~\cite{SLP,liu2021dual, liu2020jointsymbol}, the precoding  signal 
$\bm x_\ell$ is designed to shape the received signals at CUs approximately equal to the desired symbols, i.e.,  
\begin{equation}
    \bm h_k^H \bm x_\ell \approx s_{k,\ell}, ~~~\forall~\ell \in {\mathscr{L}}, ~\forall~k \in {\cal K}_U. 
\end{equation}
It is worth noting that SLP is performed on a per-symbol basis and it can exploit the instant symbol $s_{k,\ell}$ to optimize the transmit signal, thereby achieving better SEP performance than the traditional block-level precoding. It is shown in~\cite{shaomj} that given the desired symbol $s_{k,\ell}$ and the observation  in~\eqref{eq:slp_model}, the SEP at CUs, denoted as 
\begin{equation} \label{eq:sep_def}
	{\sf SEP}_{k,\ell}\triangleq {\rm Pr} \left\{ {\sf dec}(y_{k,\ell}) \neq s_{k,\ell}~|~ s_{k,\ell} \in {\cal S} \right\},
\end{equation}
 can be upper bounded by
\begin{equation} \label{eq:sep_upper}
{\sf SEP}_{k,\ell} \leq 2Q \left(\frac{\beta_{k,\ell}}{\tilde{\sigma}_k/\sqrt{2}} \sin\frac{\pi}{M} \right),
\end{equation} 
where $\beta_{k,\ell} = \Re\{ \bm h_k^H \bm x_\ell s_{k,\ell}^*\} - |\Im\{ \bm h_k^H \bm x_\ell  s_{k,\ell}^* \}| \cot(\pi/M)$. 
Therefore,  the SEP performance at the CUs can be guaranteed by imposing an upper bound on the right-hand side of~\eqref{eq:sep_upper}, i.e.
\begin{equation} \label{eq:CU_QoS}
2Q \left(\frac{\beta_{k,\ell}}{\tilde{\sigma}_k/\sqrt{2}} \sin\frac{\pi}{M} \right) \leq \epsilon_{k}, \quad \forall~k\in {\cal K}_U, \ell\in \mathscr{L}.
\end{equation} 
We will adopt~\eqref{eq:CU_QoS} as the communication performance for CUs.

\subsection{Covert and Secrecy Performance Metric}
To prevent the targets from intercepting the symbol information $s_{k,\ell}$, PLS techniques have been widely adopted in the literature to ensure information secrecy. However, as discussed in the Introduction, PLS is fundamentally an information-theoretic approach and faces practical limitations, particularly in achieving covert communication. To overcome these limitations and enable covert communication, we leverage the principle of SLP to shape the ISAC signals into noise at the malicious targets. Specifically, we impose the following noise-shaping constraint: 
\begin{equation} \label{eq:distr_constr}
\bm a_t^H (\theta_k) \bm x_\ell \sim {\cal CN}(0, |d_k|^2), \quad \forall~k\in {\cal K}_T, ~\forall~\ell \in \mathscr{L}
\end{equation}
for some \( d_k \in \mathbb{C} \), \( \forall~k\in {\cal K}_T \), where \( d_k \) should be jointly designed with \( \bm x \). This constraint enforces the received signals at the targets to follow a circularly symmetric complex Gaussian distribution, thereby eliminating recognizable modulation patterns such as PSK or QAM. Notably, while the primary purpose of the noise-shaping constraint is to enable covert communication, it also inherently serves as a form of Gaussian jamming against unintended targets.

\subsection{Problem Formulation}
Given the sensing metric in~\eqref{eq:radar_sinr}, SEP metric in~\eqref{eq:CU_QoS} and the noise-shaping metric in~\eqref{eq:distr_constr}, the ISCC problem under PSK constellation is formulated as
\begin{subequations}\label{eq:main_prob}
\begin{align}
	\hspace{-10cm}	\max_{\bm x, \bm d } & ~~ \min_{k \in {\cal K}_T} ~~   \gamma_k(\bm x)  \label{eq:main_prob_a}\\
	{\rm s.t.} & ~~  Q \left(\frac{\beta_{k,\ell}}{\tilde{\sigma}_k/\sqrt{2}} \sin\frac{\pi}{M} \right) \leq \frac{\epsilon_k}{2}, ~~\forall~k\in {\cal K}_U, \ell\in \mathscr{L}, \label{eq:main_prob_b}\\
	&~ ~ \bm a_t^H (\theta_k) \bm x_\ell \overset{\rm approx}{\sim} {\cal CN}(0, |d_k|^2) ,  \forall k\in {\cal K}_T,\ell\in \mathscr{L},\label{eq:main_prob_c} \\
	&~ ~ \| \bm x\|^2 \leq P, \label{eq:main_prob_d}
\end{align}
\end{subequations}
where $\overset{\rm approx}{\sim}$ means approximately follows the specified distribution. Note that due to \eqref{eq:main_prob_d}, the $\bm a_t^H (\theta_k) \bm x_\ell$ has  a finite support and thus   the noise-shaping constraint  holds approximately. 

By substituting~\eqref{eq:snr_radar} into \eqref{eq:main_prob} and after some manipulation of~\eqref{eq:main_prob_b}, problem~\eqref{eq:main_prob} can be equivalently expressed as
\begin{subequations}\label{eq:main_eqv}
\begin{align}
	\hspace{-10pt}	\max_{\bm x, \bm d } & ~ \min_{k \in {\cal K}_T} ~~    \bm x^H  \bm A_k^H  \bm R_k^{-1} \bm A_k \bm x   \label{eq:main_eqv_a}\\
	{\rm s.t.} & ~  \Re\{ {\bm h}_k^H \bm x_\ell \tilde{s}_{k,\ell}\} \geq {\mu}_k, ~~\forall~k  \in {\cal K}_U, ~\forall~\ell\in {\mathscr{L}},  \label{eq:main_eqv_b1}\\
& ~ \Re\{ {\bm h}_k^H \bm x_\ell \bar{s}_{k,\ell}\} \geq {\mu}_k, ~~\forall~k  \in {\cal K}_U, ~\forall~\ell\in {\mathscr{L}}, \label{eq:main_eqv_b2}\\
	&~  \bm a_t^H (\theta_k) \bm x_\ell \overset{\rm approx}{\sim}  {\cal CN}(0, |d_k|^2) ,  \forall k\in {\cal K}_T,\ell\in \mathscr{L}, \label{eq:main_eqv_c} \\
	&~  \| \bm x\|^2 \leq P, \label{eq:main_eqv_d}
\end{align}
\end{subequations}
where $\tilde{s}_{k,\ell}=s_{k,\ell}^*(\sin(\pi/M) + \mathfrak{j} \cos(\pi/M))$ and $\bar{s}_{k,\ell}=s_{k,\ell}^*(\sin(\pi/M) - \mathfrak{j} \cos(\pi/M))$ and $ \mu_k =  Q^{-1}(\epsilon_{k}/2) \tilde{\sigma}_k/\sqrt{2}$.

There are two main challenges in solving problem~\eqref{eq:main_eqv}. First, the noise-shaping constraint~\eqref{eq:main_eqv_c} cannot be easily expressed in an explicit form, making it difficult to develop an efficient optimization algorithm. Second, the objective function in~\eqref{eq:main_eqv_a} is nonconvex with respect to \( \bm x \). In the next section, we propose a tractable approximate solution to address problem~\eqref{eq:main_eqv}.

\section{The Proposed Method}\label{sec:method}

We begin by examining constraint~\eqref{eq:main_eqv_c}. To transform it into a more tractable form, we adopt a sampling-based approach and consider the following alternative formulation of~\eqref{eq:main_eqv_c}:
\begin{equation} \label{eq:confusion_inequality}
\frac{1}{L}\sum_{\ell=1}^L |  \bm a_t^H (\theta_k) \bm x_\ell - d_k  \cdot {u_{k,\ell}}  |^2 \leq \delta_k, ~~\forall~k\in {\cal K}_T,
\end{equation}
where \( \delta_k > 0 \) is a pre-specified tolerance, and \( \bm u_k = [u_{k,1}, \ldots, u_{k,L}]^T \in \mathbb{C}^L \) is a random sequence (independent of \( s_{k,\ell} \)), sampled from a standard complex Gaussian distribution, i.e., \( \bm u_k \sim {\cal CN}(\bm 0, \bm I_L) \). The inequalities in~\eqref{eq:confusion_inequality} ensure that the empirical average of the received signals at the targets closely approximates a Gaussian random sequence. Ideally, as \( \delta_k \) approaches zero, the received signals at the targets become indistinguishable from ideal Gaussian noise---independent of \( s_{k,\ell} \)---thereby preventing any information leakage.

By substituting~\eqref{eq:confusion_inequality} into \eqref{eq:main_eqv}, the main problem can be equivalently expressed as follows:
\begin{subequations}\label{eq:main_problem}
\begin{align}
	\hspace{-10pt}	\max_{\bm x, \bm d } & ~~ \min_{k \in {\cal K}_T} ~~    \bm x^H  \bm A_k^H  \bm R_k^{-1} \bm A_k \bm x   \label{eq:nmain_eqv_a}\\
	{\rm s.t.} &~ ~ 	\frac{1}{L}\sum_{\ell=1}^L |  \bm a_t^H (\theta_k) \bm x_\ell - d_k  \cdot {u_{k,\ell}}  |^2 \leq \delta_k, ~\forall~k\in {\cal K}_T, \label{eq:nmain_eqv_b} \\
	& ~~ \eqref{eq:main_eqv_b1}, \ \eqref{eq:main_eqv_b2}~ {\rm and}~ \eqref{eq:main_eqv_d}~{\rm satisfied.}
\end{align}
\end{subequations}

Next, considering the nonconvex objective in~\eqref{eq:nmain_eqv_a}, we employ the majorization minimization (MM) approach to handle it. Denote $\phi_k(\bm x) \triangleq   \bm x^H  \bm A_k^H  \bm R_k^{-1} \bm A_k \bm x$. By~\cite[Lemma 1]{8239836}, we can minorize $\phi_k(\bm x)$  by a concave (quadratic) function, given  as
\begin{equation} \label{eq:lowerbound}
	\begin{aligned}
		\phi_k(\bm x)  & \geq 	\tilde{\phi}_k(\bm x; \bar{\bm x}), ~~\forall\,\bm x, ~\bar{\bm x}, \\
		\phi_k(\bar{\bm x}) & =   	\tilde{\phi}_k(\bar{\bm x}; \bar{\bm x}),
	\end{aligned}
\end{equation}
where 
\begin{equation}
    \begin{aligned}
	& \tilde{\phi}_k(\bm x; \bar{\bm x})   \triangleq  - 2 \bm x^H \bar{\bm M}_k \bm x + 2 \Re\{ \bar{\bm m}_k^H\bm x\} +   \varphi_k, \\
	& \varphi_k = 2 \bar{\bm x}^H \bar{\bm M}_k  \bar{\bm x}  + \phi_k(\bar{\bm x}),~~ \bar{\bm m}_k =  \bm A_k^H \bar{\bm R}_k^{-1} \bm A_k \bar{\bm x}, \\
	&\bar{\bm M}_k = \sum_{c\in {\cal C}(k)} \varsigma_k^{-2}  \varsigma_c^2 \bm A_c^H  \bar{\bm R}_k^{-1} \bm A_k  {\bar{\bm x}} {\bar{\bm x}}^H  \bm A_k^H \bar{\bm R}_k^{-1} \bm A_c.
\end{aligned}
\end{equation}

Given an initial point $\bm x^{0}$,  the MM method iteratively solves the following convex surrogate problem:
\begin{subequations} \label{eq:MM_main}
	\begin{align}
		\min_{\bm x, \bm d } & ~~ \max_{k \in {\cal K}_T} ~~    -\tilde{\phi}_k(\bm x;  {\bm x}^{t}) \label{eq:MM_main_a} \\
		{\rm s.t.} & ~~  \eqref{eq:main_eqv_b1}, \ \eqref{eq:main_eqv_b2},\ \eqref{eq:main_eqv_d}~ {\rm and}~\eqref{eq:nmain_eqv_b}~{\rm satisfied},
	\end{align}
\end{subequations}
for $t=0,1,\ldots$, until some  stopping criterion is satisfied.

Since problem~\eqref{eq:MM_main} is convex, it can be optimally solved using standard solvers such as CVX. However, repeatedly invoking a general-purpose solver can result in substantial computational overhead. To address this, we focus on developing a low-complexity solution method for problem~\eqref{eq:MM_main}. The core of our approach is the proximal distance algorithm~\cite{li2020proximal}.

\subsection{Efficient Solution to Problem~\eqref{eq:MM_main} by PDA}

Problem~\eqref{eq:MM_main} can be equivalently rewritten as
\begin{subequations}\label{eq:mm_main_epi}
	\begin{align}
		\min_{\bm x, \bm d, \xi} & ~  \xi \label{eq:mm_main_epi_a}\\
		{\rm s.t.} & ~   \tilde{\phi}_k(\bm x) + \xi   \geq 0, ~k\in {\cal K}_T \label{eq:mm_main_epi_b}\\
		& ~ \Re\{ {\bm h}_k^H \bm x_\ell \tilde{s}_{k,\ell}\} \geq {\mu}_k, ~~\forall~k  \in {\cal K}_U, ~\forall~\ell\in {\mathscr{L}},  \label{eq:mm_main_epi_c1}\\
		& ~ \Re\{ {\bm h}_k^H \bm x_\ell \bar{s}_{k,\ell}\} \geq {\mu}_k, ~~\forall~k  \in {\cal K}_U, ~\forall~\ell\in {\mathscr{L}}, \label{eq:mm_main_epi_c2}\\
		&~ ~ \frac{1}{L}\sum_{\ell=1}^L |  \bm a_t^H (\theta_k) \bm x_\ell - d_k  \cdot u_{k,\ell}  |^2 \leq \delta_k, ~\forall k\in {\cal K}_T \label{eq:mm_main_epi_d} \\
		&~ ~ \| \bm x\|^2 \leq P. \label{eq:mm_main_epi_e}
	\end{align}
\end{subequations} 
For ease of exposition, denote $\bm y = \{ \bm x, \bm d,\xi\}$ and the set $ {\cal C}_i = \{ \bm y ~|~ \mbox{the $i$-th constraint in~\eqref{eq:mm_main_epi} is satisfied}\}, ~i=1,\ldots, N$ with $N= 2K_T+2K_UL+ 1$. With this notation, problem~\eqref{eq:mm_main_epi}  can be compactly written as
\begin{equation} \label{eq:PDA}
	\begin{aligned}
		\min_{\bm y } & ~ \xi \\
		{\rm s.t.} & ~ \bm y \in {\cal C}_i, ~~i=1,\ldots,N.
	\end{aligned}
\end{equation}

The key idea of the PDA  is to convert problem~\eqref{eq:PDA} into an unconstrained form by introducing a penalty term that enforces feasibility through distance minimization, and then iteratively solve the penalized problem using proximal mapping. Specifically, consider the following penalized version of problem~\eqref{eq:PDA}:
\begin{equation}\label{eq:PDA2}
	\begin{split}
		\min_{\bm y} \quad
		& \textstyle f(\bm y) \triangleq \xi + \frac{\rho}{2N} \sum_{i=1}^N \kappa(\bm y,{\cal C}_{i})^2,
	\end{split}
\end{equation}
where $\kappa(\bm y,{\cal C}_{i}) \triangleq \min_{\bm z \in {\cal C}_{i}}\|\bm y - \bm z\|_2$ denotes the distance from $\bm y$ to the set ${\cal C}_{i}$, and $\rho > 0$ is a penalty parameter.  It is known that problems~\eqref{eq:PDA} and~\eqref{eq:PDA2} become equivalent as  $\rho \to \infty$.

Furthermore, for any $\bar{\bm y}$, the distance can be upper bounded as
\begin{equation} \label{eq:dis_inequality}
	\kappa(\bm y,{\cal C}_{i}) \le \|\bm y - {\Pi}_{{\cal C}_{i}}(\bar{\bm y})\|_2 
\end{equation}
where ${\Pi}_{{\cal C}_{i}}({\bar{\bm y}}) = \arg\min_{\bm y \in {\cal C}_i} \| \bm y - \bar{\bm y}\|^2$ denotes the Euclidean projection of $\bar{\bm y}$ onto ${\cal C}_{i}$. Using this inequality, the objective function  $f(\bm y)$ in~\eqref{eq:PDA2} can be majorized as
\begin{equation}\label{eq:f_major}
	\begin{aligned}
		f(\bm y) \leq  & \xi + \frac{\rho}{2N}\sum_{i=1}^N \|\bm y - {\Pi}_{{\cal C}_{i}}(\bar{\bm y})\|_2^2,\\
		= &  \xi + \frac{\rho}{2} \| \bm y - \frac{1}{N} \sum_{i=1}^N {\Pi}_{{\cal C}_{i}}(\bar{\bm y})  \|_2^2 + \text{const}.
	\end{aligned}
\end{equation}
for any $\bar{\bm y}$, where ``const.'' denotes a constant independent of $\bm y$, and the equality holds when $\bm y = \bar{\bm y}$.  Based on this majorization, the penalized problem~\eqref{eq:PDA2} can be solved by starting from an initial (possibly infeasible) point $\bm y^0$ and iteratively performing  the following two steps:
\begin{equation} \label{eq:PAD_2steps}
	\begin{cases}
		\text{Step 1:} & \bm z^k =  \frac{1}{N} \sum_{i=1}^N {\Pi}_{{\cal C}_{i}}({\bm y}^k), \\
		\text{Step 2:} & \bm y^{k+1} = \arg\min_{\bm y} \xi + \frac{\rho}{2} \| \bm y - \bm z^k \|_2^2
	\end{cases}
\end{equation}
for $k=0,1,\ldots$ until convergence.

In this iterative process, the first step computes the projections of \( \bm y^k \) onto each constraint set \( {\cal C}_i \), and then averages the results. The second step performs a proximal update of \( \xi \) around this averaged point. Since \( f(\bm y) \) is convex, the convergence of this scheme to an optimal solution of problem~\eqref{eq:PDA2} is guaranteed~\cite{li2020proximal}.

Importantly, the efficiency of PDA hinges on the ease of computing the projection and proximal mapping steps. As we will show in Sec.~\ref{sec:proj_prox}, for the considered problem~\eqref{eq:mm_main_epi}, both steps admit closed- or semi-closed-form solutions, enabling a highly efficient implementation.

\begin{rmk}
	To further speed up convergence of PDA, one can incorporate Nesterov's acceleration strategy into~\eqref{eq:PAD_2steps}. Specifically, instead of directly using the current iterate \(\bm y^k\) for projection in Step 1, we introduce an extrapolated point \(\tilde{\bm y}^k = \bm y^k + \zeta_k(\bm y^k - \bm y^{k-1})\), where \(\zeta_k\) is the momentum parameter. The extrapolated point \(\bm v^k\) is then used in place of \(\bm y^k\) in Step 1 of~\eqref{eq:PAD_2steps}, leading to the following update:
	\begin{equation}
\begin{cases}
		\text{Step 1:} & \bm z^k = \frac{1}{N} \sum_{i=1}^N {\Pi}_{{\cal C}_{i}}(\tilde{\bm y}^k), \\
		\text{Step 2:} & \bm y^{k+1} = \arg\min_{\bm y} \xi + \frac{\rho}{2} \| \bm y - \bm z^k \|_2^2.
	\end{cases}	    
	\end{equation}
	The momentum parameter \(\zeta_k\) can be chosen using standard schemes such as \(\zeta_k = \frac{k-1}{k+\alpha}\) for some \(\alpha \geq 2\). This accelerated version of PDA can achieve faster convergence in practice while retaining the simplicity and structure of the original algorithm.
\end{rmk}
\begin{rmk}
	The penalty parameter \( \rho \) must be chosen with care. A large \( \rho \) may render problem~\eqref{eq:PDA} ill-conditioned, potentially slowing down convergence. Conversely, a small \( \rho \) may result in a poor approximation of the original constrained problem. A practical approach is to adopt a homotopy strategy, where \( \rho \) is initialized with a small value and gradually increased every few iterations. This method offers a balanced trade-off between convergence speed and solution accuracy.
	
\end{rmk}

\subsection{Projection and Proximal Mapping} \label{sec:proj_prox}

\subsubsection{Projection onto \eqref{eq:mm_main_epi_b}}

Each constraint in~\eqref{eq:mm_main_epi_b} introduces a projection problem, which is  given by
\begin{equation} \label{eq:Projection_e}
	\begin{aligned}
		\min_{\xi, \bm x} & ~ \| \bm x- \bar{\bm x}  \|^2  +  |\xi - \bar{\xi}|^2\\
		{\rm s.t.} & ~    \tilde{\phi}_k(\bm x) + \xi   \geq 0. 
	\end{aligned}
\end{equation}
Problem~\eqref{eq:Projection_e} is a quadratically constrained quadratic problem (QCQP), which can be solved by the Lagrangian method, and the solution is given by
\begin{equation} \label{eq:17b}
	\begin{aligned}
		\bm x^\star & =\begin{cases}
			\bar{\bm x},  &  \text{if~} \,\tilde{\phi}_k(\bar{\bm x}) + \bar{\xi}   \geq 0,\\
			(\bm I + 2\lambda \bar{\bm M}_k)^{-1}(\bar{\bm x}+\lambda \bar{\bm m}_k), & \text{otherwise,}\\
		\end{cases} \\
		\xi^\star& =\begin{cases}
			\bar{\xi},  & \text{if~} \,\tilde{\phi}_k(\bar{\bm x}) + \bar{\xi}   \geq 0,\\
			\bar{  \xi} - \frac{\lambda}{2}, & \text{otherwise,}\\
		\end{cases}
	\end{aligned}
\end{equation}
where  $\lambda$ is the optimal Lagrangian multiplier that can be determined by bisection search.

\subsubsection{Projection onto \eqref{eq:mm_main_epi_c1} and \eqref{eq:mm_main_epi_c2}}
Both projections can be written into the following optimization problem:
\begin{equation}\label{eq:proj_1_1}
	\begin{aligned}
		\min_{\bm z \in \mathbb{C}^{N_T}} & ~ \| \bm z -  \bar{\bm x}_{\ell}\|^2 \\
		{\rm s.t.}& ~  \Re\{ \tilde{\bm h}_k^H \bm z\} \geq \mu_k.
	\end{aligned}
\end{equation}
where $\tilde{\bm h}_k^H = \bm h_k^H \tilde{s}_{k,\ell}$ for \eqref{eq:mm_main_epi_c1} and $\tilde{\bm h}_k^H = \bm h_k^H \bar{s}_{k,\ell}$ for \eqref{eq:mm_main_epi_c2}.

As shown in Appendix~\ref{appendix:project_1}, the optimal solution of~\eqref{eq:proj_1_1} is given by
 \begin{equation} \label{eq:17c}
	\bm z^\star = { \bar{\bm x}_\ell} +  \frac{\left(\mu_k  - \Re\{ \tilde{\bm h}_k^H \bar{\bm x}_\ell\} \right)^+}{ \| \tilde{\bm h}_k\|^2}  \tilde{\bm h}_k
\end{equation} 
where $(\cdot)^+ \triangleq \max\{0, \cdot\}$.

\subsubsection{Projection onto \eqref{eq:mm_main_epi_d}}
Each constraint in~\eqref{eq:mm_main_epi_d} introduces a projection problem, which is  given by
\begin{equation}
	\begin{aligned}
		\min_{ d_k, \bm x} & ~ \sum_{\ell=1}^L\| \bm x_\ell - \bar{\bm x}_\ell \|^2  +  | d_k - \bar{d}_k|^2\\
		{\rm s.t.} & ~  \sum_{\ell=1}^L | \bm a_t^H (\theta_k) \bm x_\ell -   d_k \cdot u_{k,\ell}  |^2 \leq L \delta_k.
	\end{aligned}
\end{equation}
Denote $\bm z = [\bm x_1^T, \ldots, \bm x_L^T, d_k]^T$. The above problem can be written as
\begin{equation}
	\begin{aligned}
		\min_{  \bm z} & ~ \| \bm z - \bar{\bm z} \|^2 \\
		{\rm s.t.} & ~ \| \bm B_k \bm z  \|^2 \leq L \delta_k
	\end{aligned}
\end{equation}
where $\bm B_k \triangleq  [\bm I \otimes \bm a_t^H (\theta_k), ~ - \bm u_k]$. This is a QCQP, whose solution is given by
\begin{equation}
	\bm z^\star = (\bm I + \lambda^\star \bm B_k^H \bm B_k)^{-1} \bar{\bm z} \label{b_k},
\end{equation}
where $\lambda^\star \in \mathbb{R}_+$ is the optimal Lagrangian multiplier, which can be determined by bisection search.

\subsubsection{Projection onto \eqref{eq:mm_main_epi_e}}
The projection problem is given by
\begin{equation}\label{eq:proj_1}
	\begin{aligned}
		\min_{\bm z \in \mathbb{C}^{N_T \times L}} & ~ \| \bm z - { \bar{\bm x}}\|^2 \\
		{\rm s.t.}& ~  \left\|\bm{z}\right\|^2 \leq P
	\end{aligned}
\end{equation}
Similarly, this is a QCQP, whose solution is given by
\begin{equation} \label{eq:17e}
	\bm z^\star = \begin{cases}
		\bar{\bm x}, & \mbox{if $\|\bar{\bm x}\|^2 \leq P$,} \\
		\frac{\sqrt{P}}{\| \bar{\bm x} \| }\bar{\bm x}, & {\rm otherwise.}
	\end{cases}
\end{equation}

\subsubsection{Proximal Mapping}
The proximal mapping in Step 2 requires solving the following problem:
\begin{equation}
	\begin{aligned}
		\min_{\bm x, \bm d, \xi} & ~  \xi + \frac{\rho}{2} (|   \xi - \bar{\xi}^k |^2 + \| \bm x - \bar{\bm x}^k \|^2+\| \bm d - \bar{\bm d}^k \|^2)    \\
	\end{aligned}
\end{equation}
where $(\bar{\xi}^k, \bar{\bm x}^k, \bar{\bm d}^k)$ represents the average $\bm z^k$ in Step 1 of~\eqref{eq:PAD_2steps}. It is easy to see that the optimal solution is given by 
\begin{equation}\label{eq:prox_map_solution}
	\bm x^\star= \bar{\bm x}^k, \quad \bm d^\star = \bar{\bm d}^k, \quad \xi^\star = \bar{\xi}^k - {1}/{\rho}
\end{equation}

By combining the MM and PDA update steps described above, we develop an MM-PDA iterative algorithm for solving the ISCC problem~\eqref{eq:main_problem}, as summarized in Algorithm~\ref{alg:2}.

\begin{algorithm}
	\caption{MM-PDA for ISCC Problem~\eqref{eq:main_problem}} \label{alg:2}
	\begin{algorithmic}[1]
		\State \textbf{Input} $\rho_{max}$, $\mu_k$,  $\bm y^0 = (\bm x^0, \bm d^0, \xi^0)$, $K$
		\State \textbf{Set} $\epsilon>0$, $\rho > 0$, $\kappa > 1$, $k = 0$ and $t=0$	
		\Repeat
		\State Construct the surrogate problem \eqref{eq:MM_main} at $\bm y^t$
		\Repeat
		\State Compute projection onto \eqref{eq:mm_main_epi_b} by \eqref{eq:17b}
		\State Compute projection onto \eqref{eq:mm_main_epi_c1}, \eqref{eq:mm_main_epi_c2}  by \eqref{eq:17c}
		\State Compute projection onto \eqref{eq:mm_main_epi_d} by \eqref{b_k}
		\State Compute projection onto \eqref{eq:mm_main_epi_e} by \eqref{eq:17e}
		\State Compute proximal mapping by \eqref{eq:prox_map_solution}
		\State $k\leftarrow k+1$
		\State Update $\rho$ = min$\{\kappa\rho, \rho_{max}\}$ every $K$ iterations
		\Until{PDA stopping criterion is satisfied}
		\State $\bm y^{t+1} \leftarrow \bar{\bm y}^k$
		\State $t\leftarrow t+1$, $k\leftarrow 0$
		\Until{$\| \xi^t- \xi^{t-1} \| <\epsilon$}
		\State \textbf{Output}: $\bm x^t$
	\end{algorithmic}
\end{algorithm}

\subsection{Convergence and Complexity Analysis}
The proposed MM-PDA algorithm employs two nested iterative schemes: an outer MM loop and an inner PDA loop. In each MM iteration, a tight convex surrogate of the original non-convex problem is constructed, and the resulting convex subproblem is solved using the PDA. According to the convergence results of the PDA framework~\cite{li2020proximal}, the inner loop converges to the global optimum of each convex MM subproblem. By combining this with the convergence property of MM methods~\cite{hunter2004tutorial}, it follows that if each MM subproblem is solved exactly---which is ensured here by the convergence of the inner PDA---the overall MM-PDA algorithm converges to a stationary point of the original non-convex problem. This nested convergence guarantee ensures both theoretical soundness and numerical stability in practice.

Next, let us analyze the complexity of MM-PDA.
MM-PDA uses an outer MM loop (say $m$ iterations) and an inner proximal-distance loop (say $p$ iterations). The dominant costs per inner iteration are:
1) SEP (communication) projections: there are $O(K_U L)$ scalar constraints. Each projection needs one inner product $\bm{h}_k^H\bm{x}_\ell$ and a rank-1 correction, giving $O(K_U L N_t)$.
2) Noise-shaping projections (per target): each requires solving a QCQP with $\bm B_k \in \mathbb{C}^{L \times (LN_t+1)}$. Using the SVD of $\bm B_k \bm B_k^H$ and bisection search gives $O(L^2 + L N_t)$ per target, i.e., $O(K_T(L^2 + L N_t))$.
3) Radar surrogate projections (per target): each update involves $(\mathbf{I}+2\lambda\bar{\mathbf{M}}_k)^{-1}$ applied to a vector of size $LN_t$. $\bar{\mathbf{M}}_k$ depends on both the target and clutter steering matrices, and the complexity scales with both $N_t$ and $N_r$. By exploiting the Kronecker/block structure, the cost reduces to $O\big((K_T+K_C)(N_t^2 L^2 + N_r^2 L^2)\big)$.
4) Power-ball projection and proximal update: linear in $LN_t$. 
Hence the overall complexity is $O\!\Big(mp\,[\,K_U L N_t \;+\; K_T(L^2+L N_t) \;+\; (K_T+K_C)(N_t^2 L^2+N_r^2 L^2)]\Big).$
This is significantly lighter than generic interior-point solvers, whose cost scales super-cubic in the variable dimension $(\sim LN_t)$.

\section{Generalization to QAM Modulation} \label{sec:QAM}
In this section, we extend the symbol-level ISCC design to support QAM constellations. Due to the non-uniform structure of QAM, the original communication performance metrics must be appropriately reformulated. Specifically, the received baseband discrete-time signal at the \(k\)-th CU remains as defined in \eqref{eq:slp_model}. Let the desired symbol \(s_{k,\ell}\) for user \(k\) be drawn from a \(4M^2\)-QAM constellation set:
\begin{equation}
	\mathcal{S}=\left\{s_{R}+\mathfrak{j}s_{I} \mid s_{R}, s_{I} \in \{ \pm 1, \pm 3, \cdots, \pm(2M - 1) \} \right\}
\end{equation}
where \(M\) is a positive integer determining the constellation order, and \(s_{R}\) and \(s_{I}\) denote the real and imaginary parts of the symbol, respectively. 

We jointly optimize the transmit signal sequence \(\{\boldsymbol{x}_{\ell}\}_{\ell\in \mathscr{L}}\) and a complex-valued dynamic range vector \(\boldsymbol{\tau} = [\tau_{1}, \cdots, \tau_{K}]^{\mathrm{T}}\), with \(\tau_{k} = \tau_{k}^{R} + \mathfrak{j} \tau_{k}^{I}\),  $\tau_{k}^{R}\geq 0$ and $\tau_{k}^{I}\geq 0$, to establish the following approximation:
\begin{equation} \label{eq:QAMCU}
	\boldsymbol{h}_{k}^{H}\boldsymbol{x}_{\ell} \approx \tau_{k}^{R} \mathfrak{R}(s_{k,\ell}) + \mathfrak{j} \tau_{k}^{I} \mathfrak{I}(s_{k,\ell}).
\end{equation}
Here, \(\tau_{k}^{R}\) and \(\tau_{k}^{I}\) represent the dynamic range (i.e., half-spacing) of the real and imaginary parts of the received signal for user \(k\), respectively.

According to~\cite{SLP}, for the \(4M^2\)-QAM constellation, the SEP in~\eqref{eq:sep_upper} holds if the following condition is satisfied:
\begin{equation} \label{eq:vecform_csep}
	-\boldsymbol{\tau} + \boldsymbol{a}_{\ell} \leq_{c} \boldsymbol{H} \boldsymbol{x}_{\ell} - \boldsymbol{\tau} \diamond \boldsymbol{s}_{\ell} \leq_{c} \boldsymbol{\tau} - \boldsymbol{c}_{\ell},
\end{equation}
where $\bm H = [\bm h_1, \ldots, \bm h_K]^H$, $\boldsymbol{x} \leq_{c} \boldsymbol{y}$ denotes element-wise comparison over both the real and imaginary parts, i.e., $\Re(\boldsymbol{x}) \leq \Re(\boldsymbol{y})$ and $\Im(\boldsymbol{x}) \leq \Im(\boldsymbol{y})$. The symbol-wise scaling operation $\boldsymbol{\tau} \diamond \boldsymbol{s}_{\ell}$ is defined as
\begin{equation}
    [\boldsymbol{\tau} \diamond \boldsymbol{s}_{\ell}]_k = \tau_k^{R} \Re(s_{k,\ell}) + \mathfrak{j} \tau_k^{I} \Im(s_{k,\ell}).
\end{equation}
The complex vectors $\boldsymbol{a}_{\ell} = [a_{1,\ell}^R + \mathfrak{j} a_{1,\ell}^I  , \ldots, a_{K,\ell}^R + \mathfrak{j} a_{K,\ell}^I ]^T \in \mathbb{C}^K$ and $\boldsymbol{c}_{\ell} = [c_{1,\ell}^R + \mathfrak{j} c_{1,\ell}^I  , \ldots, c_{K,\ell}^R + \mathfrak{j} c_{K,\ell}^I ]^T \in \mathbb{C}^K$ are calculated as
\begin{equation}\label{eq:a_c_def}
	\begin{aligned}
		a_{k,\ell}^{R}&=
		\begin{cases}
			\alpha_{k}, & \vert\mathfrak{R}(s_{k,\ell})\vert<2M - 1\\
			\beta_{k}, & \mathfrak{R}(s_{k,\ell})=2M - 1\\
			-\infty, & \mathfrak{R}(s_{k,\ell})=-2M + 1
		\end{cases}\\
		c_{k,\ell}^{R}&=
		\begin{cases}
			\alpha_{k}, & \vert\mathfrak{R}(s_{k,\ell})\vert<2M - 1\\
			-\infty, & \mathfrak{R}(s_{k,\ell})=2M - 1\\
			\beta_{k}, & \mathfrak{R}(s_{k,\ell})=-2M + 1
		\end{cases}
	\end{aligned}
\end{equation}
\begin{equation} \label{eq:alpha_k}
	\alpha_{k}=\frac{\tilde{\sigma}_{k}}{\sqrt{2}}Q^{-1}\left(\frac{1 - \sqrt{1 - \epsilon_{k}}}{2}\right),\beta_{k}=\frac{\tilde{\sigma}_{k}}{\sqrt{2}}Q^{-1}\left(1 - \sqrt{1 - \epsilon_{k}}\right)
\end{equation}
and $a_{k,\ell}^I$ and $c_{k,\ell}^I$ are calculated similarly as $a_{k,\ell}^R$ and $c_{k,\ell}^R$ in~\eqref{eq:a_c_def} with $R$ and $\mathfrak{R}$ replaced by $I$ and $\mathfrak{I}$, respectively.

By substituting the SEP constraints in~\eqref{eq:mm_main_epi_c1}-\eqref{eq:mm_main_epi_c2} with \eqref{eq:vecform_csep}, the optimization problem for the QAM constellation can be similarly expressed  as follows:
\begin{subequations}\label{eq:qam_main_epi}
	\begin{align}
		\min_{\bm x, \bm v, \xi, \boldsymbol{\tau}} & ~ \xi \label{eq:qam_main_epi_a}\\
		{\rm s.t.} & ~ -\!\boldsymbol{\tau} + \boldsymbol{a}_{\ell}\leq_{c}\boldsymbol{\upsilon}_{\ell} \! - \! \boldsymbol{\tau}\diamond \boldsymbol{s}_{\ell}\leq_{c}\boldsymbol{\tau} - \boldsymbol{c}_{\ell}, ~\forall \ell\in {\mathscr{L}} \label{eq:qam_main_epi_c}\\
		& ~\bm \tau\geq_c \bm 0, \label{eq:qam_main_epi_c1} \\
		& ~ \boldsymbol{\upsilon}_{\ell} = \boldsymbol{H} \boldsymbol{x}_{\ell}, ~\forall \ell\in {\mathscr{L}} \label{eq:qam_main_epi_d}\\
		& ~ \eqref{eq:mm_main_epi_b},~ \eqref{eq:mm_main_epi_d}~ \text{and}~ \eqref{eq:mm_main_epi_e}.  \label{eq:qam_main_epi_e}
	\end{align}
\end{subequations}

The QAM-based optimization problem can still be tackled using the PDA framework developed for the PSK case. However, compared with the PSK setting, extending Algorithm~\ref{alg:2} to QAM modulation introduces two main challenges: 
(i) the non-constant modulus of QAM symbols leads to per-symbol amplitude box constraints in both the in-phase and quadrature components, complicating the feasible set; 
(ii) all transmit vectors within a CPI share a common constellation scaling factor, which couples the SEP constraints across timeslots and makes the projection step significantly more involved, especially for large CPIs. 
In the following, we focus on the projection steps associated with the newly introduced constraints in \eqref{eq:qam_main_epi_c} and \eqref{eq:qam_main_epi_d}; the remaining constraints' projection and the proximal mapping can be handled as in the PSK case.

\subsubsection{Projection onto Constraint (\ref{eq:qam_main_epi_c}) and \eqref{eq:qam_main_epi_c1}}
The projection problem associated with the constraints~\eqref{eq:qam_main_epi_c} and \eqref{eq:qam_main_epi_c1} is  given by
\begin{subequations}\label{eq:proj_un_d}
	\begin{align}
		\min_{\boldsymbol{\upsilon}, \boldsymbol{\tau}} & ~ \|\boldsymbol{\tau} - \bar{\boldsymbol{\tau}} \|^2 + \sum_{\ell=1}^{\mathscr{L}} \|\boldsymbol{\upsilon}_\ell - \bar{\boldsymbol{\upsilon}}_\ell \|^2 \label{eq:proj_un_d_a}\\
		\rm{s.t.} & ~ -\boldsymbol{\tau} + \boldsymbol{a}_\ell \leq_c \boldsymbol{\upsilon}_\ell - \boldsymbol{\tau} \diamond \boldsymbol{s}_\ell \leq_c \boldsymbol{\tau} - \boldsymbol{c}_\ell, ~\forall~\ell \in {\mathscr{L}}  \label{eq:proj_un_d_b} \\
		& ~ \bm \tau \geq_c \bm 0 \label{eq:proj_un_d_c} 
	\end{align}
\end{subequations}

Remarkably, the projection problem (\ref{eq:proj_un_d}) exhibits complete separability across both user indices \(k = 1,\cdots,K\) and complex dimensions (real/imaginary parts). This key property enables decomposition into \(2K\) independent subproblems, each sharing the following unified structure:
\begin{subequations}
\begin{align}
	\min_{\{\upsilon_\ell\}_{\ell=1}^{\mathscr{L}}, \tau} & ~ |\tau - \bar{\tau}|^2 + \sum_{\ell=1}^{\mathscr{L}} |\upsilon_\ell - \bar{\upsilon}_\ell|^2 \label{eq:proj_un_d_seq_a} \\
	\rm{s.t.} & ~ -\tau + a_\ell \leq_c \upsilon_\ell - \tau \diamond s_\ell \leq_c \tau - c_\ell, ~\forall \ell \label{eq:proj_un_d_seq_b}\\
	&~  \tau \geq_c 0, \label{eq:proj_un_d_seq_c}
\end{align}
\end{subequations}
where \(\tau\), \(\upsilon_\ell\), \(s_\ell\), \(a_\ell\), and \(c_\ell\) represent the corresponding components of vectors \(\boldsymbol{\tau}\), \(\boldsymbol{\upsilon}_\ell\), \(\boldsymbol{s}_\ell\), \(\boldsymbol{a}_\ell\), and \(\boldsymbol{c}_\ell\) with identical row indices. The real and imaginary parts remain decoupled in both objective and constraints, allowing separate treatment of \((\Re\{\upsilon_\ell\},\Re\{\tau\})\) and \((\Im\{\upsilon_\ell\},\Im\{\tau\})\). The subsequent analysis focuses on the real part solution, with the imaginary part following identical procedures.

For notational simplicity, we reuse symbols \(\tau\), \(\upsilon_\ell\), \(\bar{\tau}\), \(\bar{\upsilon}_\ell\), \(a_\ell\), \(c_\ell\), and \(s_\ell\) to denote their real parts. The simplified problem becomes:
\begin{subequations} \label{eq:proj_un_d_eq_seq}
	\begin{align}
	\min_{\{\upsilon_\ell\}_{\ell=1}^{\mathscr{L}}, \tau} & ~ (\tau - \bar{\tau})^2 + \sum_{\ell=1}^{\mathscr{L}} (\upsilon_\ell - \bar{\upsilon}_\ell)^2 \label{eq:proj_un_d_eq_seq_a}\\
	\rm{s.t.} & ~ (s_\ell-1)\tau + a_\ell \leq \upsilon_\ell \leq (s_\ell+1)\tau - c_\ell,~~\forall \ell \label{eq:proj_un_d_eq_seq_b}\\
	& \tau \geq 0. 
\end{align}
\end{subequations}

For any fixed \(\tau\), the optimal \(\upsilon_\ell^\star(\tau)\) admits a closed-form solution:
\begin{equation} \label{eq:opt_v_solu}
   \upsilon_\ell^\star(\tau) = 
\begin{cases}
	(s_\ell-1)\tau + a_\ell, & \bar{\upsilon}_\ell \leq l_\ell(\tau) \\
	\bar{\upsilon}_\ell, & l_\ell(\tau) \leq \bar{\upsilon}_\ell \leq u_\ell(\tau) \\
	(s_\ell+1)\tau - c_\ell, & \bar{\upsilon}_\ell \geq u_\ell(\tau)
\end{cases} 
\end{equation}
where \(l_\ell(\tau) = (s_\ell-1)\tau + a_\ell\) and \(u_\ell(\tau) = (s_\ell+1)\tau - c_\ell\). 
The piecewise nature of \(\upsilon_\ell^\star(\tau)\) is determined by two critical points:
\begin{enumerate}
	\item Intersection with the lower bound: 
	\begin{equation}
	    \tau_{\ell,1} = (\bar{\upsilon}_\ell - a_\ell)/(s_\ell-1),~\forall \ell \in \mathscr{L}.
	\end{equation}
	\item Intersection with the upper bound:  \begin{equation}
	    \tau_{\ell,2} = (\bar{\upsilon}_\ell + c_\ell)/(s_\ell+1),~\forall \ell \in \mathscr{L}.
	\end{equation} 
\end{enumerate}
All of these critical points partition $\tau$ into distinct regions, within each of which the optimal $\upsilon_\ell^\star(\tau)$ takes the form given in~\eqref{eq:opt_v_solu}. By evaluating the objective function in~\eqref{eq:proj_un_d_eq_seq_a} across these regions and selecting the $\tau$ that yields the minimum value, we obtain the optimal pair $(\tau^\star, \upsilon_\ell^\star(\tau^\star))$. The detailed procedure for identifying $\tau^\star$ is provided in Appendix~\ref{appendix_search_tau}.

\subsubsection{Projection onto the constraint \eqref{eq:qam_main_epi_d}}
Each constraint in~\eqref{eq:qam_main_epi_d} introduces a projection problem, which is  given by
\begin{equation}
	\begin{aligned}
		\min_{\boldsymbol{\upsilon}_\ell, \boldsymbol{x}_\ell} & ~ \|\boldsymbol{\upsilon}_\ell - \bar{\boldsymbol{\upsilon}}_\ell \|^2 + \|\boldsymbol{x}_\ell - \bar{\boldsymbol{x}}_\ell \|^2  \\
		\mathrm{s.t.} & ~ \boldsymbol{\upsilon}_\ell = \boldsymbol{H} \boldsymbol{x}_\ell. 
	\end{aligned}
\end{equation}

By substituting the equality constraint into the objective, we get an unconstrained least squares problem with respect to $\bm x_\ell$ and thus  the optimal $\bm x^\star_\ell$ is given by
\begin{equation}
	\boldsymbol{x}_{\ell}^\star = (\boldsymbol{H}^{H}\boldsymbol{H} + \boldsymbol{I})^{-1}(\overline{\boldsymbol{x}}_{\ell} + \boldsymbol{H}^{H}\overline{\boldsymbol{\upsilon}}_{\ell}).
\end{equation}
and the optimal $\boldsymbol{\upsilon}_{\ell}^\star = \boldsymbol{H}\boldsymbol{x}_{\ell}^\star$.

 \section{Robust ISCC Optimization under Imperfect CSI} \label{sec:robust}
 \subsection{Imperfect CSI Model}
 In the preceding sections, the ISCC optimization is developed under the assumption of perfect CSI, which is idealized in practice. In this section, we extend the framework to the case of imperfect CSI for both CUs and malicious targets. Specifically, we adopt bounded uncertainty models as follows. For each CU,
 \begin{equation} \label{eq:csi_erro_CU}
 	\bm h_k \in {\cal H}_k \triangleq \{ \hat{\bm h}_k + \Delta \bm h_k~|~ \|\Delta \bm h_k\|_2\le \varepsilon_{U,k} \},~ \forall k\in {\cal K}_U,
 \end{equation}
 where $\hat{\bm h}_k$ is the estimated CSI and $\varepsilon_{U,k}>0$ specifies the radius of the uncertainty ball. For each target,
 \begin{equation} \label{eq:csi_error_target}
 	\theta_k  \in {\bm  \Theta}_k \triangleq \{ \hat{\theta}_k + \Delta \theta_k~|~  |\Delta \theta_k| \leq  {\varepsilon}_{T,k} \}, \quad \forall k\in {\cal K}_T,
 \end{equation}
 where $\hat{\theta}_k$ is the estimated angle and  ${\varepsilon}_{T,k}>0$ denotes the angular deviation bound. 
 
 \subsection{Robust Transmit Design for PSK Constellation}
 Under the above uncertainty sets, the robust ISCC design is formulated as
 \begin{subequations}\label{eq:robust_psk}
 	\begin{align}
 		\max_{\bm x, \bm d } & ~ \min_{k \in {\cal K}_T} ~~ \min_{\theta_k \in \Theta_k} ~~ \bm x^H  \bm A_k^H  \bm R_k^{-1} \bm A_k \bm x   \label{eq:robust_psk_a}\\
 		{\rm s.t.} ~~
 		& \min_{\bm h_k\in {\cal H}_k}\Re\{ {\bm h}_k^H \bm x_\ell \tilde{s}_{k,\ell}\} \geq {\mu}_k, ~\forall k,\ell, \label{eq:robust_psk_b}\\
 		& \min_{\bm h_k\in {\cal H}_k}\Re\{ {\bm h}_k^H \bm x_\ell \bar{s}_{k,\ell}\} \geq {\mu}_k, ~\forall k,\ell, \label{eq:robust_psk_c}\\
 		& \min_{\theta_k \in \Theta_k} \tfrac{1}{L}\sum_{\ell=1}^L |  \bm a_t^H (\theta_k) \bm x_\ell - d_k  u_{k,\ell}  |^2 \leq \delta_k, ~\forall k, \label{eq:robust_psk_d} \\
 		& \| \bm x\|^2 \leq P.   \label{eq:robust_psk_e}
 	\end{align}
 \end{subequations}
Problem~\eqref{eq:robust_psk} aims at maximizing the worst-case ISCC performance by considering all possible $\Delta \bm h_k$ and $\Delta \theta_k$.

We first consider the SEP constraints in~\eqref{eq:robust_psk_b} and \eqref{eq:robust_psk_c}. The following result shows their exact second-order cone (SOC) reformulation. 
 \begin{proposition}\label{prop:robust_sep_eqv}
 	The robust SEP constraints in~\eqref{eq:robust_psk_b}–\eqref{eq:robust_psk_c} are equivalent to the following inequalities: for each symbol slot $\ell$, there exists a slack variable $r_\ell \ge 0$ such that
 	\begin{equation}\label{eq:robust_sep_soc}
 		\begin{aligned}
 			\left\{
 			\begin{aligned}
 				\Re\{\hat{\bm h}_k^H \bm x_\ell \tilde{s}_{k,\ell}\} &\ge \mu_k+\varepsilon_{U,k} r_\ell, && \forall k,\ell, \\
 				\Re\{\hat{\bm h}_k^H \bm x_\ell \bar{s}_{k,\ell}\} &\ge \mu_k+\varepsilon_{U,k} r_\ell, && \forall k,\ell, \\
 				\|\bm x_\ell\|_2 &\le r_\ell, && \forall \ell.
 			\end{aligned}
 			\right.
 		\end{aligned}
 	\end{equation}
 	
 \end{proposition}
 
\noindent {\it Proof.}~See Appendix~\ref{sec:proof_prop1}. \hfill $\blacksquare$

 Next, we address the target-side uncertainty. The angle set $\Theta_k=[\hat\theta_k-\varepsilon_{T,k},\,\hat\theta_k+\varepsilon_{T,k}]$ makes problem~\eqref{eq:robust_psk} a semi-infinite program. To handle this, we approximate the min/max over $\Theta_k$ by discretization:
\begin{equation}
\mathcal{G}_k \triangleq \{\theta_k^{(m)}\}_{m=1}^{M_k}\subset \Theta_k,	
\end{equation} 
 where the grid can be chosen uniformly or via Chebyshev nodes. With this discretization, the worst-case SCNR objective in~\eqref{eq:robust_psk_a} is approximated by 
 \begin{equation} \label{eq:robust_scnr_discrete}
 	\min_{\theta_k^{(m)}\in\mathcal{G}_k} \ \bm x^H \bm A_k^H(\theta_k^{(m)}) \bm R_k^{-1}\bm A_k(\theta_k^{(m)}) \bm x,
 \end{equation}
 while the covertness constraints in~\eqref{eq:robust_psk_d} are enforced for all $\theta_k^{(m)}\in\mathcal{G}_k$, i.e.,
 \begin{equation} \label{eq:robust_covert_discrete}
 	\tfrac{1}{L}\sum_{\ell=1}^L\big|\bm a_t^H(\theta_k^{(m)})\bm x_\ell - d_k\,u_{k,\ell}\big|^2 \le \delta_k,~\forall \theta_k^{(m)}\in\mathcal{G}_k,\ \forall k.
 \end{equation}

By substituting~\eqref{eq:robust_sep_soc}, \eqref{eq:robust_scnr_discrete} and \eqref{eq:robust_covert_discrete} into problem~\eqref{eq:robust_psk} and  introducing a slack variable $\xi$ to capture the worst-case SCNR objective, the robust ISCC problem becomes
	\begin{align}
		\max_{\bm x,\bm d,\bm r,\xi}\quad & \xi  \notag \\
		\text{s.t.}\quad 
		& \xi \le \bm x^H \bm A_k^H(\theta_k^{(m)}) \bm R_k^{-1} \bm A_k(\theta_k^{(m)}) \bm x,~~\forall \theta_k^{(m)}\in\mathcal{G}_k, ~\forall k,  \notag\\
		& \Re\{\hat{\bm h}_k^H \bm x_\ell \tilde{s}_{k,\ell}\} \ge \mu_k+\varepsilon_{U,k} r_\ell, ~\forall k,\ell,  \notag\\
		& \Re\{\hat{\bm h}_k^H \bm x_\ell \bar{s}_{k,\ell}\} \ge \mu_k+\varepsilon_{U,k} r_\ell,~\forall k,\ell,  \notag\\
		& \|\bm x_\ell\|_2 \le r_\ell,\ \ r_\ell\ge 0, ~\forall \ell,  \notag\\
		& \tfrac{1}{L}\sum_{\ell=1}^L |\bm a_t^H(\theta_k^{(m)})\bm x_\ell - d_k \cdot u_{k,\ell}|^2 \le \delta_k, ~\forall \theta_k^{(m)}\in\mathcal{G}_k, \forall k, \notag\\
		& \|\bm x\|_2^2 \le P. \label{eq:robust_psk_final_obj}
			\end{align} 

 As in the perfect-CSI case, problem~\eqref{eq:robust_psk_final_obj} can be handled under the MM-PDA framework: at iteration $t$, each quadratic term 
 $\bm x^H \bm A_k^H(\theta_k^{(m)}) \bm R_k^{-1}\bm A_k(\theta_k^{(m)}) \bm x$ 
 is replaced with its concave surrogate $\tilde{\phi}_{k,m}(\bm x;\bm x^{t})$, yielding convex constraints 
 $\xi \le \tilde{\phi}_{k,m}(\bm x;\bm x^{t})$. 
 The resulting subproblem is a convex quadratic program and can be solved efficiently by the proposed PDA solver.

\subsection{The Robust Transmit Design for QAM Constellation}
Since the QAM case differs from the PSK case in the SEP constraint only, we focus on developing a robust counterpart of the SEP constraint in~\eqref{eq:vecform_csep}  for the QAM constellation (the remaining targets-related constraints are the same as~\eqref{eq:robust_psk_final_obj}), which is given as
\begin{equation} \label{eq:robust_sep_qam}
		\left\{
\begin{aligned}
\min_{\bm h_k \in {\cal H}_k}\bm h_k^H \bm x_\ell & \geq_c -\tau_k + a_{k,\ell}+ \tau_k \diamond s_{k,\ell} , ~\forall k, \ell,  \\
\max_{\bm h_k \in {\cal H}_k} \bm h_k^H \bm x_\ell & \leq_c \tau_k - c_{k,\ell} + \tau_k \diamond s_{k,\ell} ,  ~\forall k, \ell,  
\end{aligned}
\right.
\end{equation}
where the $\min$ is taken separately for real and imaginary part of $\bm h_k^H \bm x_\ell$, i.e. $\min_{\bm h_k \in {\cal H}_k}\bm h_k^H \bm x_\ell = \min_{\bm h_k \in {\cal H}_k} \Re\{\bm h_k^H \bm x_\ell\} + \mathfrak{j}  \min_{\bm h_k \in {\cal H}_k} \Im\{\bm h_k^H \bm x_\ell\} $ and similar meaning applies for the $\max$ operation.

Regarding~\eqref{eq:robust_sep_qam}, we have the following equivalent reformulation:
 \begin{proposition}\label{prop:robust_sep_QAM_eqv}
	The robust SEP constraint in~\eqref{eq:robust_sep_qam} is equivalent to the following inequalities: for each symbol slot $\ell$, there exists a slack variable $r_\ell \ge 0$ such that
	\begin{equation} \label{eq:robust_sep_qam_eqv}
		\left\{
		\begin{aligned}
			 \hat{\bm h}_k^H \bm x_\ell & \geq_c -\tau_k + a_{k,\ell}+ \tau_k \diamond s_{k,\ell}  + \varepsilon_{U,k} {\tilde{r}}_\ell, ~\forall k, \ell,  \\
			 \hat{\bm h}_k^H \bm x_\ell & \leq_c \tau_k - c_{k,\ell} + \tau_k \diamond s_{k,\ell}  - \varepsilon_{U,k} {\tilde{r}}_\ell,  ~\forall k, \ell, \\
			 \| \bm x_\ell\|_2 & \leq  r_\ell, ~\forall \ell, 
		\end{aligned}
		\right.
	\end{equation}
or more compactly as
	\begin{equation}\label{eq:robust_sep_QAM_eqv_compact}
		\left\{
		\begin{aligned}			
	& -\bm \tau + \bm a_\ell + {\bm \varepsilon}_U {\tilde{r}}_\ell  \leq_c \hat{\bm H} \bm x_\ell - \bm \tau \diamond \bm s_\ell \leq_c \bm \tau - \bm c_\ell -  {\bm \varepsilon}_U {\tilde{r}}_\ell , ~\forall \ell \\
	& \| \bm x_\ell \|_2 \leq r_\ell, ~\forall \ell  
		\end{aligned}
		\right.
	\end{equation}
	where ${\tilde{r}}_\ell =  r_\ell + \mathfrak{j} r_\ell$, $\hat{\bm H} = [\hat{\bm h}_1, \ldots, \hat{\bm h}_K]^H$,  $\bm \varepsilon_U = [\varepsilon_{U,1}, \ldots, \varepsilon_{U,K}]^T$.
\end{proposition}

\noindent The proof of Proposition~\ref{prop:robust_sep_QAM_eqv} follows the same steps as Proposition~\ref{prop:robust_sep_eqv}, by noting that 
\begin{align}
& \min_{\bm h_k \in {\cal H}_k} \Re\{\bm h_k^H \bm x_\ell\} = \Re\{\hat{\bm h}_k^H \bm x_\ell\} - \varepsilon_{U,k} r_\ell, \\
& \min_{\bm h_k \in {\cal H}_k} \Im\{\bm h_k^H \bm x_\ell\} = \Im\{\hat{\bm h}_k^H \bm x_\ell\} - \varepsilon_{U,k} r_\ell.
\end{align}
We omit the detailed proof due to the page limit.

Since the constraints in~\eqref{eq:robust_sep_QAM_eqv_compact} are convex and essentially the same as the perfect-CSI case in~\eqref{eq:vecform_csep}. Hence, the resulting  robust ISCC problem can be handled under the MM framework.

\section{Numerical Results} \label{sec:num_results}
In this section, we provide simulation results to show the effectiveness of the proposed transmit waveform design algorithms. The following settings are assumed throughout our simulations. The BS is equipped with the same number of transmit and receive antennas $N_t=N_r = N=15$.
We assume that the noise power and QoS requirements for each communication user are identical, with normalized noise power $\tilde{\sigma}^2$ = $\tilde{\sigma}^2_k$ = 1 and $\mu$ = $\mu_k$, $\forall k$.  The transmission energy is set to $P = 30$. The targets are located at $\theta_{t0}$ = $-30^{\circ}$ and $\theta_{t1}$ = $30^{\circ}$. And the CUs are located at $\theta_{k0}$ = $-25^{\circ}$ and $\theta_{k1}$ = $25^{\circ}$. The target angle and user angle are the default unless explicitly stated.

We assume that $\bm h_k$ is a slow time-varying block Rician fading channel, i.e., the channel is constant in a block but varies slowly from one block to another. Thus, the channel vector of the $k$-th user can be expressed as a combination of a deterministic strongest line-of-sight (LoS) channel vector and a multi-path scattered channel vector:
\begin{equation} \label{eq:y_c}
	\bm h_k = \sqrt\frac{v_k}{1 + v_k} \bm h_{L,k}^{LoS} + \sqrt\frac{1}{1 + v_k} \bm h_{S,k}^{NLoS},
\end{equation}
where $v_k > 0$ is the Rician factor of the $k$-th user, $\bm h_{L,k}^{LoS} = \sqrt{N_t} \mathbf {a}_{t}(\omega_{k, 0})$ is the LoS deterministic component, and $\omega_{k, 0} \in [-\frac{\pi}{2}, \frac{\pi}{2}]$ is the angle of departure (AoD) of the LoS component from the BS to user $k$. The scattering component $\bm h_{S,k}^{NLoS}$ can be expressed as 
\begin{equation}
	\bm h_{S,k}^{NLoS} = \sqrt \frac{N_t}{L_P} \sum\limits_{l=1}^{L_P} c_{k,l} \mathbf {a}_{t}(\omega_{k, 0}),
\end{equation} where $L_P$ denotes the number of propagation paths, $c_{k,l} \sim \mathcal {CN}(0, 1)$ is the complex path gain and $\omega_{k, l} \in [-\frac{\pi}{2}, \frac{\pi}{2}]$ is the AOD associated to the $(k,l)$-th propagation path.

We use ISCC to denote the proposed symbol-level ISCC design with noise-shaping constraints, and SLP to denote the conventional symbol-level precoding algorithm of~\cite{alodeh2015constructive} without noise shaping. 
In addition, traditional linear beamforming is included as a baseline and denoted by BF, where the beamformers are optimized once per coherence block to maximize the worst-case sensing SCNR under QoS constraints derived from the SEP requirements. The resulting fixed beamformers are then applied to all symbols in the block, with sensing SCNR evaluated using the same radar processing chain as in ISCC. 
For notational clarity in the experiments, user 1 is denoted as U1, malicious target 1 as T1, and the case where target 1 attempts to eavesdrop on user 1 is represented by (T1$\rightarrow$U1), and so on are defined in the same way.

	\begin{table}[!h]
	\renewcommand\arraystretch{1.2}
	\caption{Average runtime (in seconds) for different number of users; $N=8$} 
	\centering
	\begin{tabular}{c c c c c}
		\toprule
		$K$   & 2      & 3      & 4      & 5 \\
		\midrule
		CVX   & 7.2702 & 9.1661 & 10.6154 & 12.3949 \\
		PDA   & 0.8572 & 0.9361 & 1.0251  & 1.1961  \\
		\bottomrule
	\end{tabular}
	\label{tab1}
\end{table}

\begin{table}[!h]
	\renewcommand\arraystretch{1.2}
	\caption{Average runtime (in seconds) for different number of antennas; $K=2$} 
	\centering
	\begin{tabular}{c c c c c}
		\toprule
		$N$   & 4      & 8      & 16     & 32 \\
		\midrule
		CVX   & 3.4762 & 7.2702 & 9.4305 & 15.5748 \\
		PDA   & 0.5127 & 0.8572 & 1.0926 & 1.3481 \\
		\bottomrule
	\end{tabular}
	\label{tab2}
\end{table}

\subsection{Convergence Performance}
\begin{figure}[t]
		\centering
		\subfloat[Convergence of objective value.]{
			\includegraphics[width=0.45\linewidth,trim=2 10 30 30,clip]{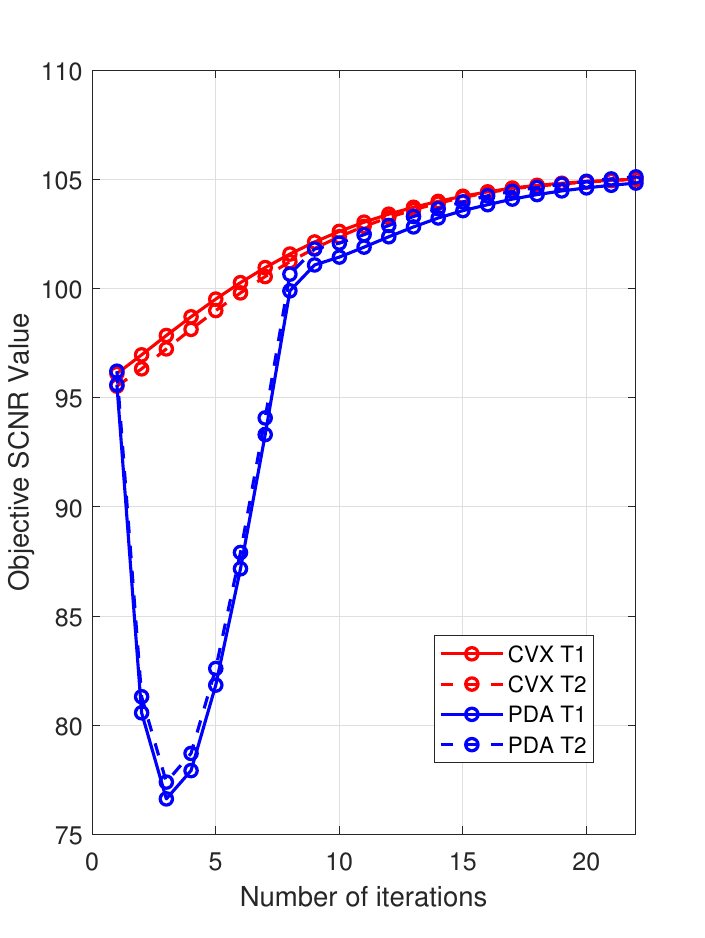}
			\label{fig:conv_obj}
		}
		\hfil
		\subfloat[Convergence of difference of successive iterates.]{
			\includegraphics[width=0.45\linewidth,trim=2 10 30 30,clip]{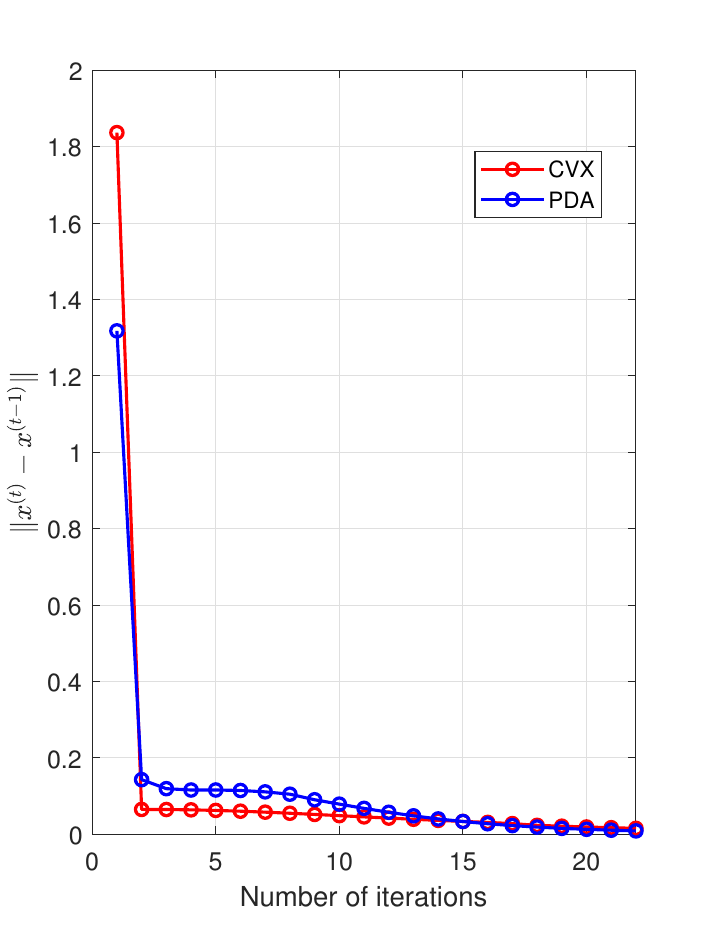}
			\label{fig:conv_diff}
		}
		\caption{Convergence performance.}
		\label{fig:conv_all}
	\end{figure}

	Fig.~\ref{fig:conv_all} illustrates the convergence of the objective value of problem \eqref{eq:main_problem} and the difference between successive iterates $\|\bm x^{t}-\bm x^{t-1}\|_2$ with respect to the MM iteration number.
	As shown in Fig.~\ref{fig:conv_all}(a), the objective curve of our algorithm first decreases due to the gradual increase of the penalty parameter $\rho$ (avoid ill-conditioning \cite{li2020proximal}), which drives the solution from the infeasible region into the feasible region. Once feasibility is ensured, the iterations follow the standard MM procedure and converge stably, achieving performance comparable to the CVX benchmark.  Meanwhile, Fig.~\ref{fig:conv_all}(b) demonstrates that the iteration error diminishes rapidly and vanishes within about $20$ iterations.
	
	Furthermore, the runtime performance of CVX and PDA is compared in Table~\ref{tab1} and Table~\ref{tab2} for different numbers of CUs and antennas. The results clearly demonstrate that the proposed PDA achieves a substantial reduction in runtime compared to CVX.

\subsection{The PSK Case}
In this section, we evaluate the performance of the proposed design under PSK modulation, with QPSK chosen as a representative example for clarity of presentation.  For a fair comparison with the conventional precoding schemes, in which the SNR threshold is $\Gamma$, we set the QoS requirement of our proposed schemes as $\mu = \tilde{\sigma} \sin (\pi/M) \sqrt{\Gamma}$.

\begin{figure}[!h]
	\vspace{-10pt}
	\centerline{\resizebox{.4\textwidth}{!}{\includegraphics{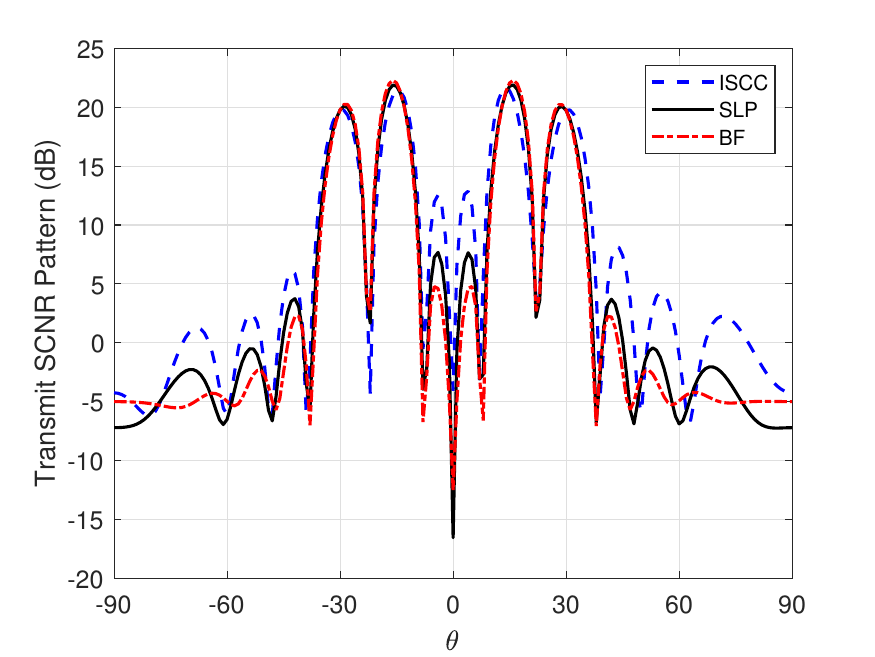}}
	}   \caption{Instantaneous transmit beampattern comparison.} \label{fig:beampattern}
\end{figure}

\begin{figure}[!h]
	\vspace{-10pt}
	\centerline{\resizebox{.4\textwidth}{!}{\includegraphics{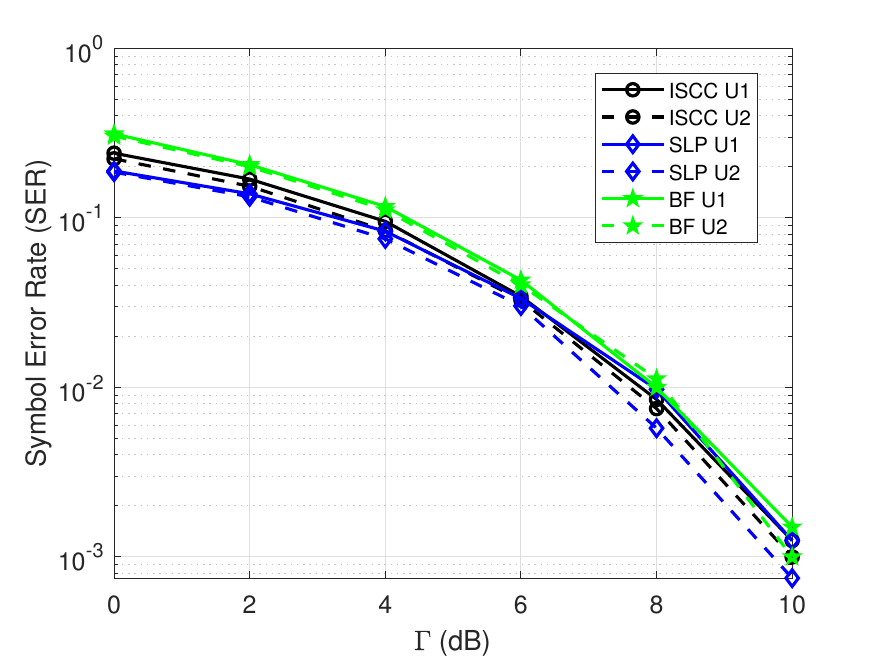}}
	}   \caption{SER of CUs vs. QoS requirements $\Gamma$ under QPSK.} \label{fig:CU_SER_QPSK}
\end{figure}
\begin{figure}[!h]
	\vspace{-10pt}
	\centerline{\resizebox{.4\textwidth}{!}{\includegraphics{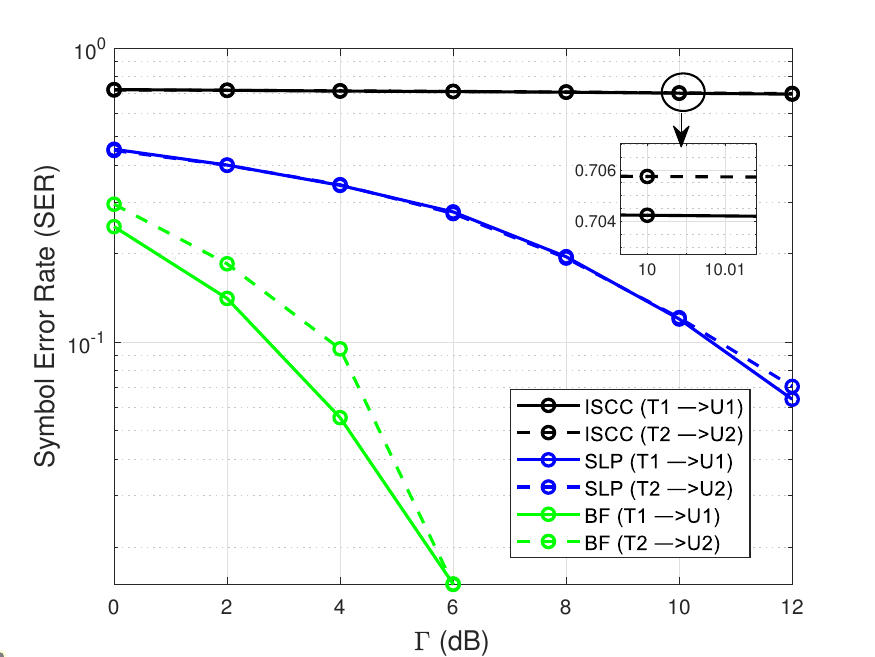}}
	}   \caption{Malicious targets' SER comparison among different schemes.} \label{fig:tar_SER}
\end{figure}
\begin{figure}[!h]
	\vspace{-10pt}
	\centerline{\resizebox{.4\textwidth}{!}{\includegraphics{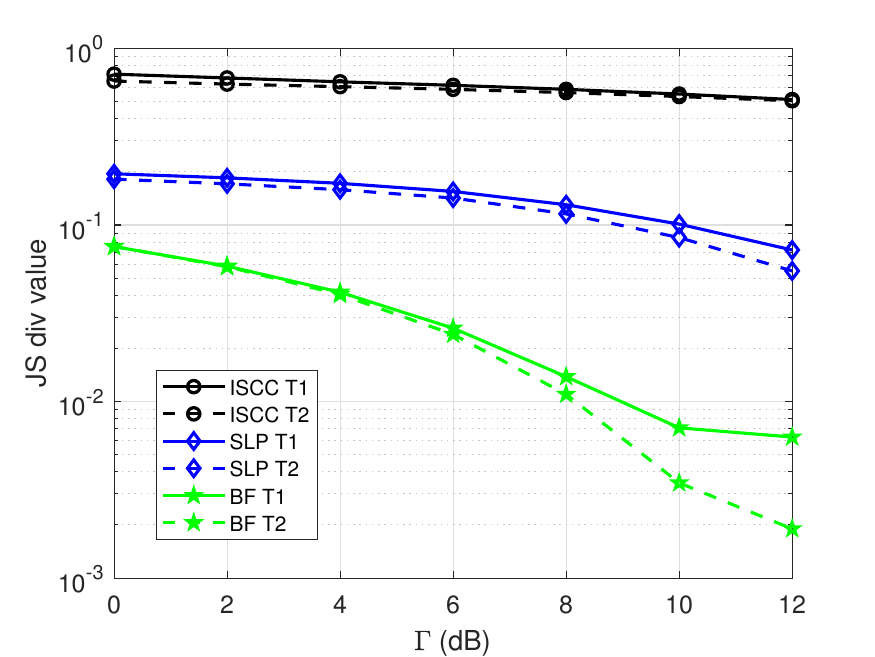}}
	}   \caption{JS divergence versus SNR threshold under QPSK.} \label{fig:jsdiv_QPSK}
\end{figure}

Fig.~\ref{fig:beampattern} illustrates the distribution of the transmit-side radar echo SCNR across spatial angles at the ISAC transmitter. Two targets are located at $\theta_{t0}$ = $-15^{\circ}$ and $\theta_{t1}$ = $15^{\circ}$, and two CUs are located at $\theta_{k0}$ = $-30^{\circ}$ and $\theta_{k1}$ = $30^{\circ}$. Comparing the patterns of ISCC, SLP, and traditional beamforming, we observe that although ISCC enforces an explicit covertness constraint, its SCNR levels at the two malicious targets and two communication users are comparable to those of SLP and beamforming without such a constraint. This indicates that the proposed design achieves negligible sensing-performance loss while providing additional covertness guarantees—highlighting the advantage of our joint optimization framework. Moreover, we should point out that due to the incorporated covertness constraint, a high SCNR in the malicious target directions does not translate into higher detectability at the malicious targets, because the ISCC design explicitly shapes the malicious target's received waveform to remain statistically close to Gaussian noise, thereby maintaining covert-communication performance. This effect will be further demonstrated in Fig.~\ref{fig:tar_SER} and Fig.~\ref{fig:jsdiv_QPSK}  through the warden’s SER and received waveform distribution.

In Fig.~\ref{fig:CU_SER_QPSK}, the average SER performance of multiple communication users under QPSK modulation is illustrated.  
It is observed that the SER consistently decreases as the QoS threshold $\Gamma$ increases. Compared to conventional beamforming, the SLP method achieves a noticeable performance gain, benefiting from its non-linear precoding design. However, the ISCC scheme, which incorporates PLS constraints, shows a slightly higher SER than SLP. This indicates that enforcing secure constraints sacrifices part of the communication efficiency, reflecting the inherent trade-off between reliability and security.

\begin{figure}[!h]
	\centerline{\resizebox{.4\textwidth}{!}{\includegraphics{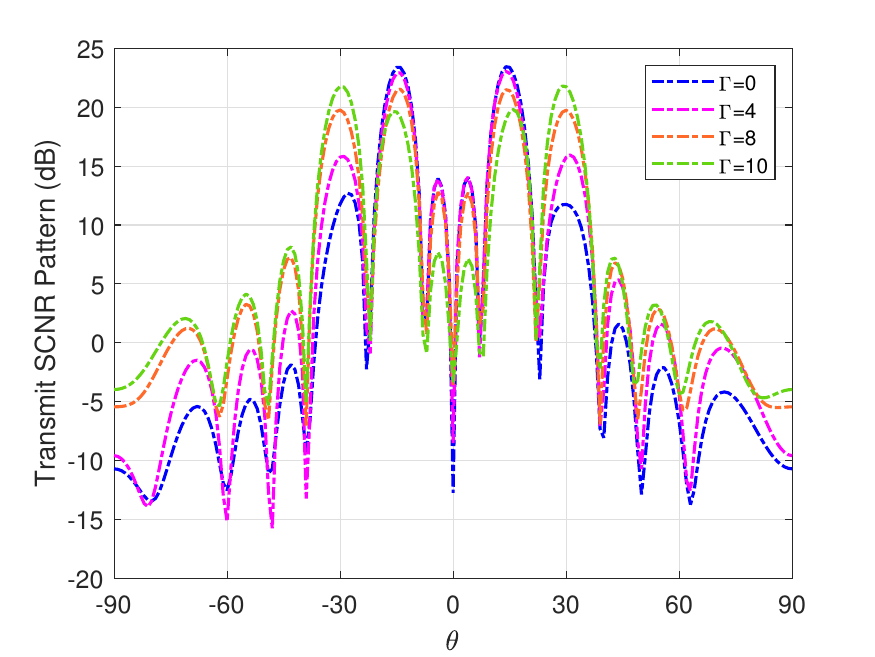}}
	}   \caption{Beampattern comparison under different $\Gamma$.} \label{fig:beamgamma}
\end{figure}

Fig.~\ref{fig:beamgamma} presents the corresponding beampatterns under different $\Gamma$ values. As $\Gamma$ increases, more transmit power is directed towards the users to satisfy stricter QoS requirements, which leads to enhanced beampattern peaks in the user directions. Consequently, the power allocated to malicious targets decreases slightly. Nevertheless, the main lobes toward sensing directions remain stable, confirming that the proposed ISCC method successfully balances sensing and communication without severely degrading the sensing capability.

\begin{figure*}[t]
	\centering
	\subfloat[Beampattern]{
		\includegraphics[width=0.3\textwidth,trim=10 2 20 15,clip]{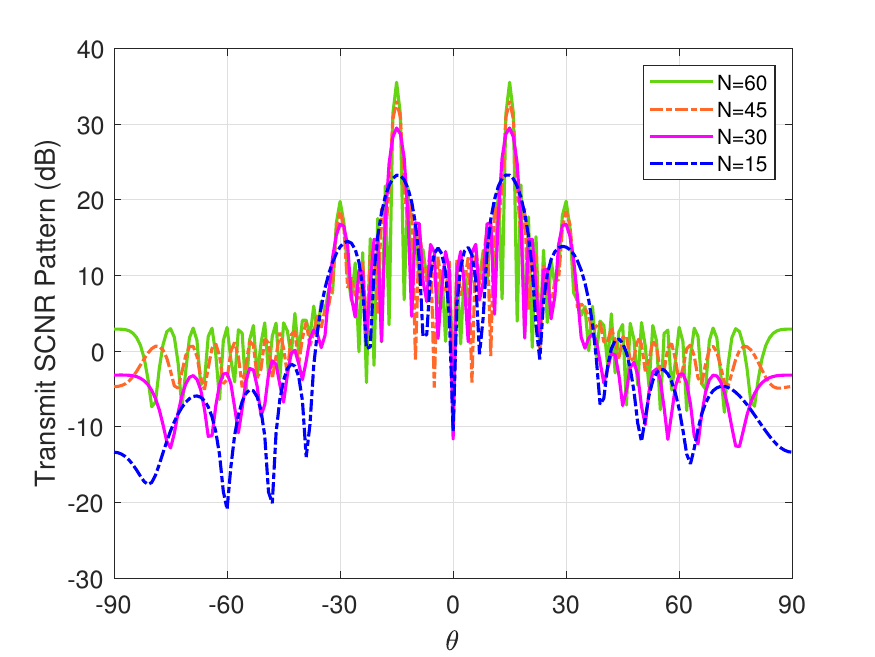}
	}\hspace{0.02\textwidth}
	\subfloat[CUs' SER]{
		\includegraphics[width=0.3\textwidth,trim=10 2 20 15,clip]{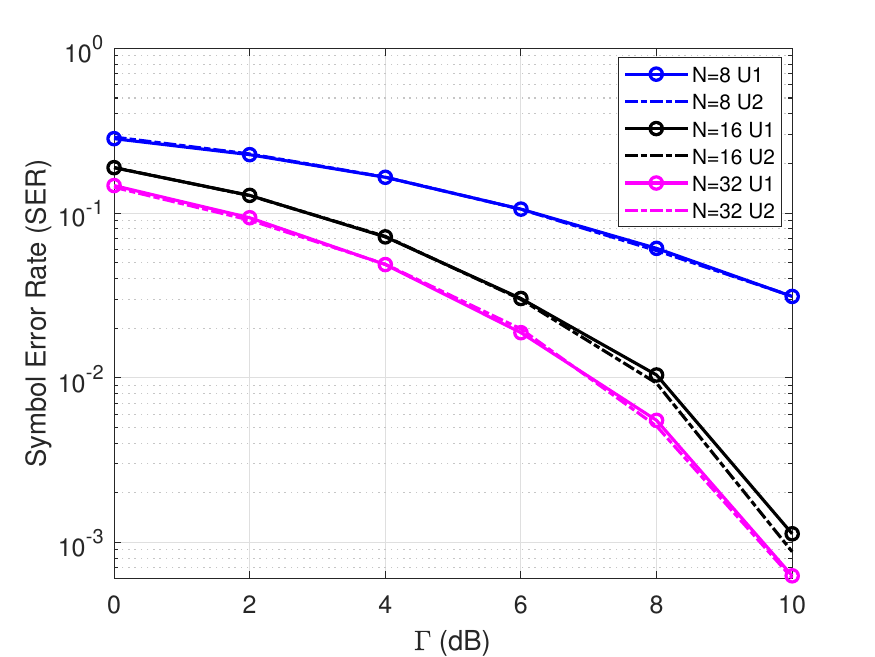}
	}\hspace{0.02\textwidth}
	\subfloat[Malicious targets' SER]{
		\includegraphics[width=0.3\textwidth,trim=10 2 20 15,clip]{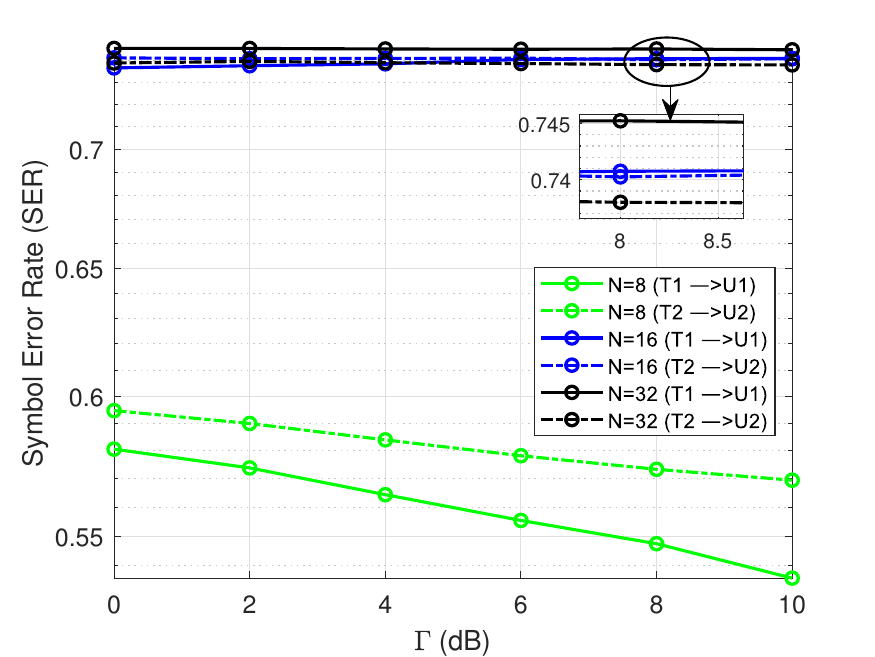}
	}
	\caption{Performance under different number of antennas.}
	\label{fig:Nantennas}
\end{figure*}

\begin{figure*}[!h]
	\centering
	\subfloat[BF]{
		\includegraphics[width=0.25\textwidth]{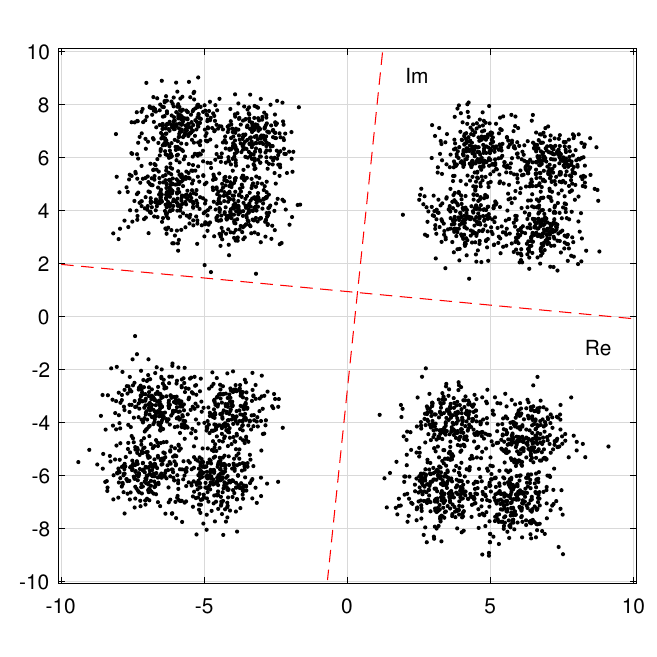}
		\label{fig:bf}
	}\hfil
	\subfloat[SLP]{
		\includegraphics[width=0.25\textwidth]{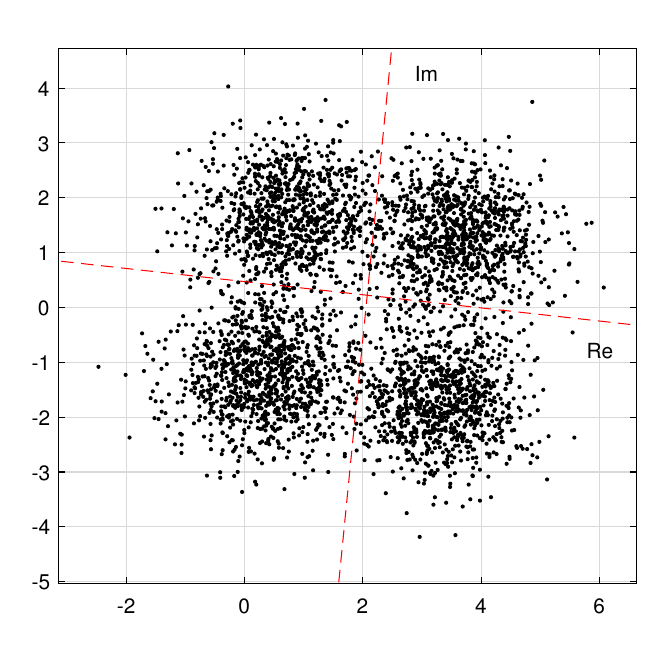}
		\label{fig:slp}
	}\hfil
	\subfloat[ISCC ($\delta=0.1$)]{
		\includegraphics[width=0.25\textwidth]{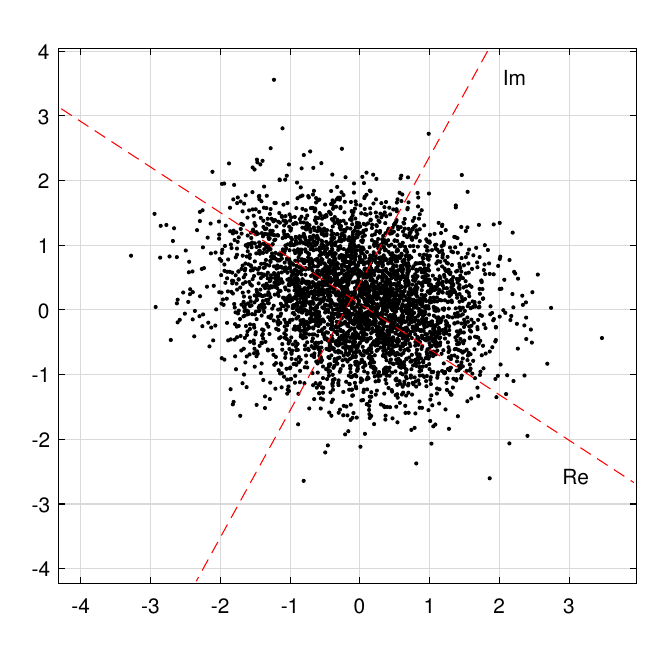}
		\label{fig:secslp}
	}\\
	\vspace{-8pt}
	\subfloat[ISCC ($\delta=10$)]{
		\includegraphics[width=0.25\textwidth,trim=10 5 10 0,clip]{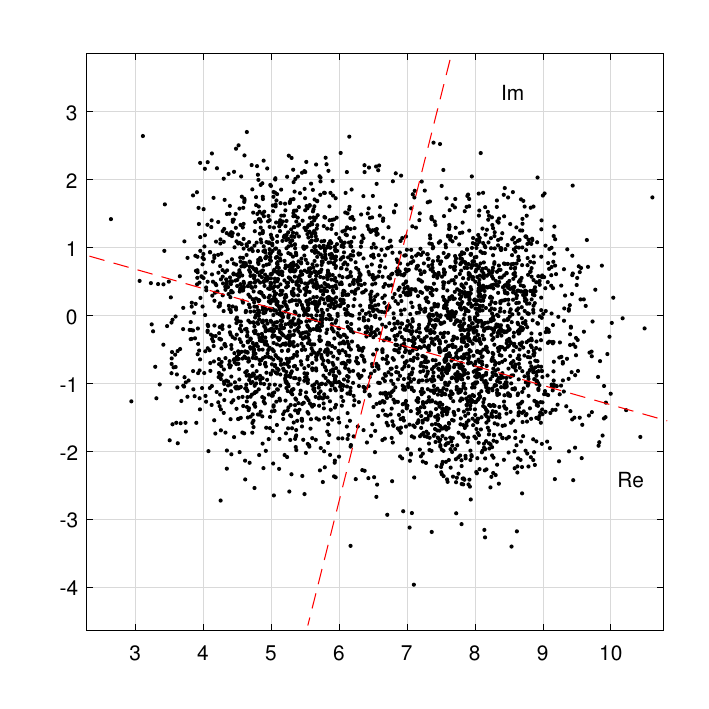}
		\label{fig:delta10}
	}\hfil
	\subfloat[ISCC ($\delta=1$)]{
		\includegraphics[width=0.25\textwidth,trim=10 5 10 0,clip]{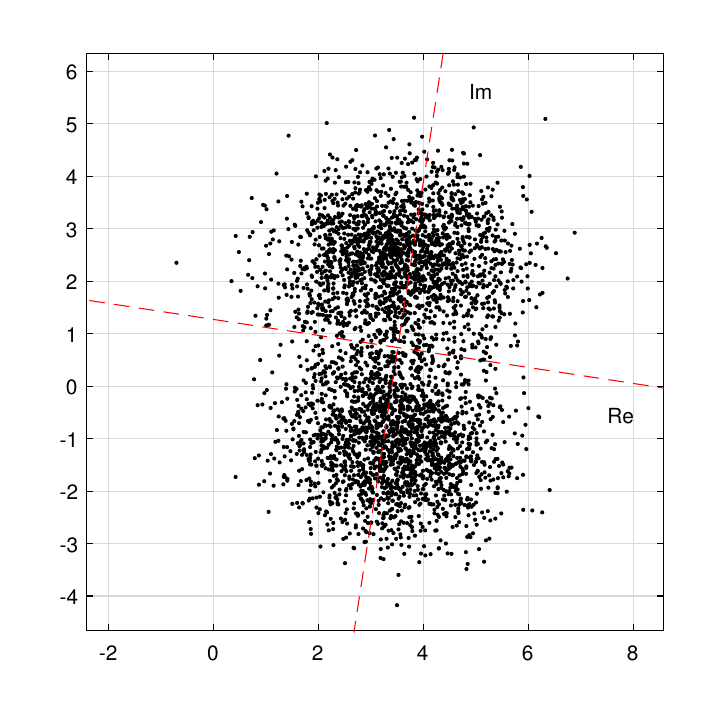}
		\label{fig:delta1}
	}\hfil
	\subfloat[ISCC ($\delta=0.1$)]{
		\includegraphics[width=0.25\textwidth]{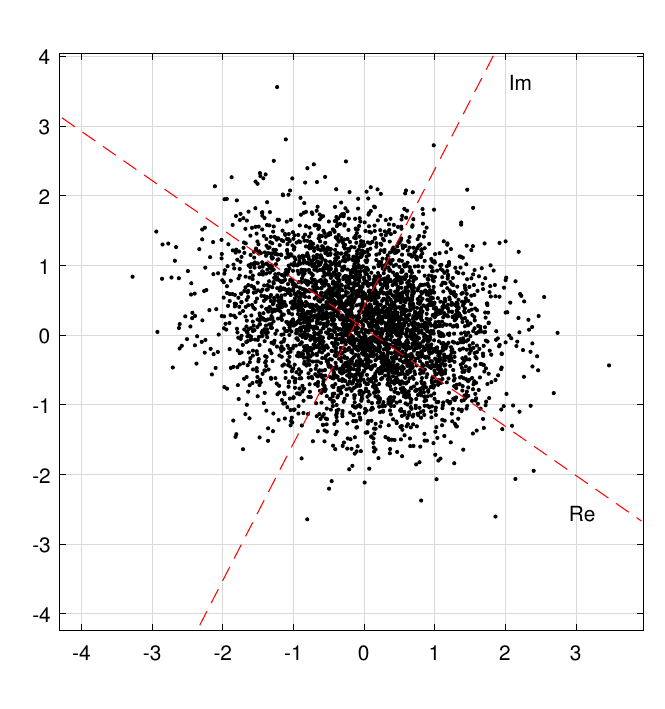}
		\label{fig:delta01}
	}
	
	\caption{The constellation of received signals at the malicious targets under QPSK modulation.}
	\label{fig:constellation_QSPK}
\end{figure*}
Fig.~\ref{fig:tar_SER} illustrates the SER performance at the malicious targets. To emphasize the worst-case security risk, the best interception case—defined as the minimum SER among all malicious targets—is selected as the representative result. As $\Gamma$ increases, the malicious targets’ SER decreases, reflecting their improved interception capability. Conventional beamforming is highly vulnerable, and even the SLP scheme fails to prevent the malicious targets’ SER from dropping significantly in high-SNR regimes, thereby posing a serious security threat. In contrast, the proposed ISCC approach consistently maintains the malicious targets’ SER at a high level across the entire $\Gamma$ range, clearly demonstrating its effectiveness in mitigating information leakage while simultaneously ensuring reliable communication and satisfactory sensing performance.

To further investigate the impact of the antenna array size, Fig.~\ref{fig:Nantennas}(a) shows the corresponding beampatterns. Larger antenna arrays generate narrower and higher mainlobes, which enhance spatial resolution and improve the SCNR at the malicious targets. The SER performance of the CUs is depicted in Fig.~\ref{fig:Nantennas}(b). As $N$ increases, the SER decreases significantly, confirming that a larger array provides more spatial degrees of freedom for interference suppression and constructive signal design. Then, Fig.~\ref{fig:Nantennas}(c) illustrates the malicious targets’ performance. Systems with larger $N$ maintain higher SER at the malicious targets, thereby strengthening covertness and effectively mitigating information leakage.

The constellation distribution of the received symbols at the targets is illustrated in Fig.~\ref{fig:constellation_QSPK}. In the upper row, the beamforming and SLP schemes both reveal QPSK-like structures after phase correction, which enables the target to detect potential communication activity. By contrast, the proposed ISCC scheme yields randomized, noise-like distributions, where no clear modulation pattern can be identified, thus effectively preventing symbol recovery. The lower row in Fig.\ref{fig:constellation_QSPK}(d)—Fig.\ref{fig:constellation_QSPK}(f)  demonstrates the effect of the  noise-shaping constraint $\delta_k$. With a large $\delta_k$ (e.g., $\delta_k=10$), residual modulation features are still visible in the constellation. As $\delta_k$ decreases, the received symbols become progressively more randomized. At $\delta_k=0.1$, the constellation fully resembles Gaussian noise, thereby concealing the modulation structure and achieving strong covertness.

Next, we perform a quantitative analysis of the received symbol distributions at both the user and the radar target. 
To evaluate the effectiveness of the proposed intrinsic covertness mechanism, we adopt the Jensen–Shannon (JS) divergence between the received symbol distributions at the communication user and at the malicious target as a quantitative metric. This choice directly reflects our covertness definition: if the warden’s received waveform is statistically indistinguishable from circularly symmetric complex Gaussian noise, its distribution will deviate significantly from that of the legitimate user’s structured constellation, yielding a high JS divergence. Unlike conventional covert metrics such as detection error probability, which require a specific hypothesis-testing model at the warden, the JS divergence provides an implementation-agnostic measure that captures the degree to which modulation structure is eliminated at the warden, in line with the proposed symbol- and noise-shaping design.
As shown in Fig.~\ref{fig:jsdiv_QPSK}, the ISCC method consistently achieves higher JS divergence than SLP and beamforming, indicating greater dissimilarity between the distributions observed by CUs and malicious targets. This structural difference obscures the modulation patterns at the malicious targets, thereby validating the covertness performance of ISCC.

\subsection{The QAM Case}
In this section, we evaluate the performance of the proposed design under QAM modulation, with 16QAM chosen as a representative example for clarity of presentation. We set the QoS requirement of the proposed schemes as $\mu = Q^{-1}(\epsilon/2)\tilde{\sigma}/\sqrt{2}$, where $\epsilon$ denotes the upper bound of the SEP at the CUs.

\begin{figure}[!h]
	\vspace{-10pt}
	\centerline{\resizebox{.4\textwidth}{!}{\includegraphics{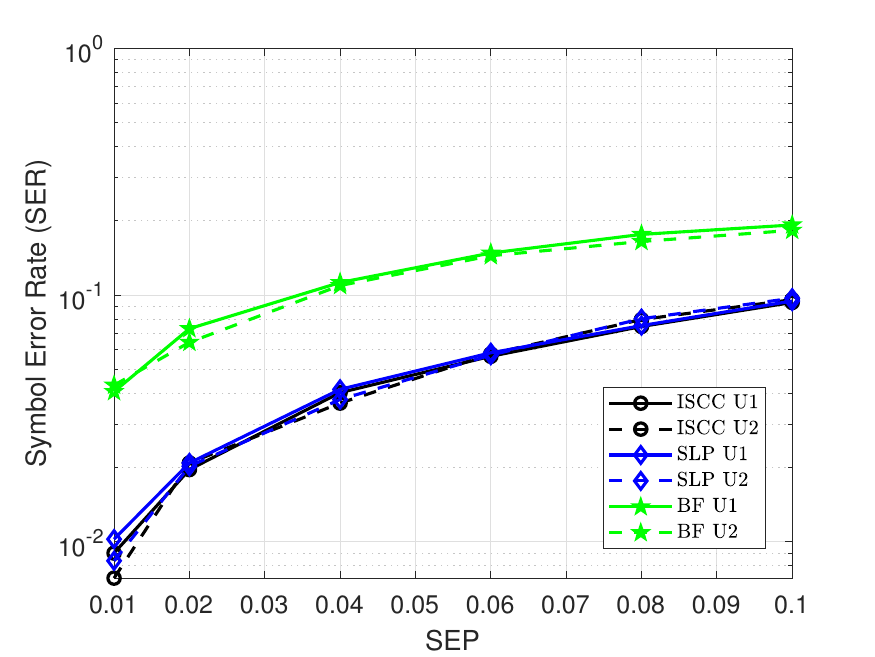}}
	}   \caption{SER of CUs versus SEP threshold under 16QAM.} \label{fig:CU_SER_QAM}
\end{figure}

Fig.~\ref{fig:CU_SER_QAM} depicts the relationship between the SER and the SEP threshold for the three considered methods. As shown, both the ISCC and SLP schemes successfully meet the SEP threshold constraints, with their SER performance remaining within the specified limits. In contrast, the conventional beamforming approach, which is not explicitly designed to satisfy SEP requirements, fails to adhere to the threshold constraints.
\begin{figure}[!h]
	\vspace{-10pt}
	\centerline{\resizebox{.4\textwidth}{!}{\includegraphics{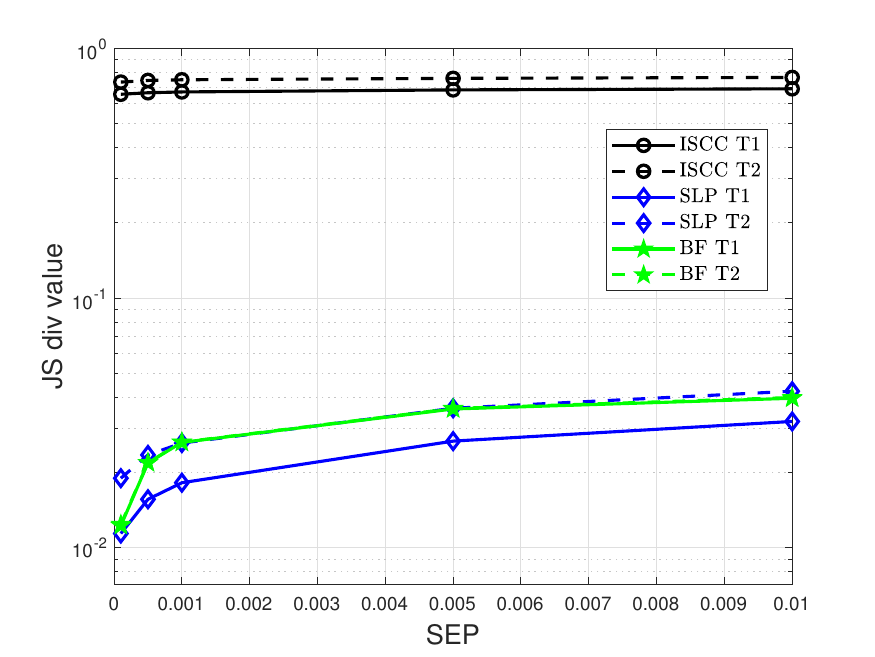}}
	}   \caption{JS divergence value versus SEP threshold under 16QAM.} \label{fig:jsdiv_QAM}
\end{figure}

\addtocounter{figure}{1}
\begin{figure}[!h]
	\centerline{\resizebox{.43\textwidth}{!}{\includegraphics{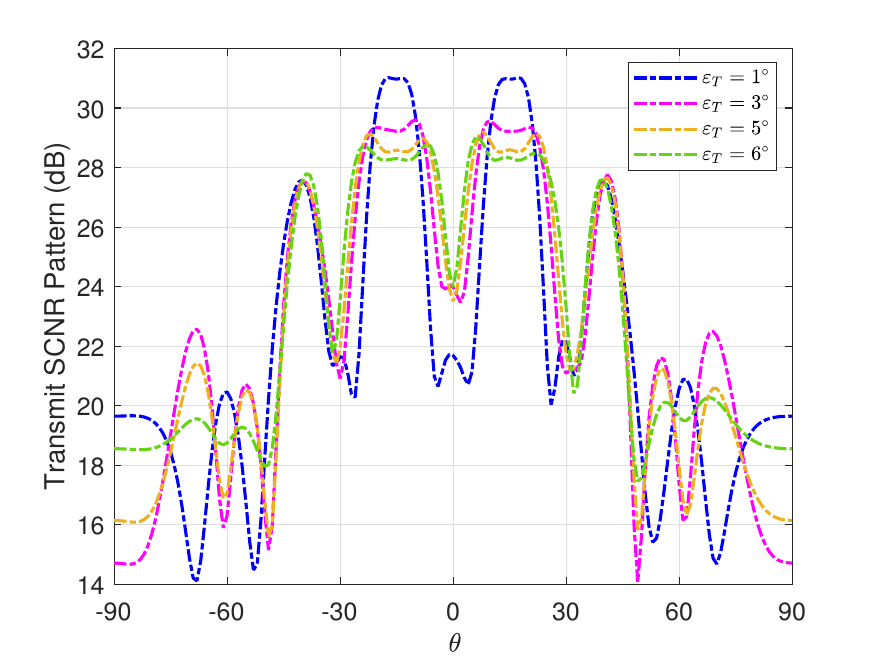}}}
	\caption{Beampattern comparison under different ${\varepsilon}_{T}$.}
	\label{fig:beamet}
\end{figure}

As shown in Fig.~\ref{fig:jsdiv_QAM}, the JS divergence value for ISCC is significantly higher compared to both the SLP and beamforming. This indicates a substantial difference in the statistical structure between the symbol distributions received by the malicious target and the legitimate user under the ISCC scheme. Such a distributional disparity makes it difficult for the malicious target to accurately identify the modulation scheme of the user's communication through statistical analysis.

\addtocounter{figure}{-2}
\begin{figure*}[t]
	\centering
	\subfloat[BF]{
		\includegraphics[width=0.25\textwidth]{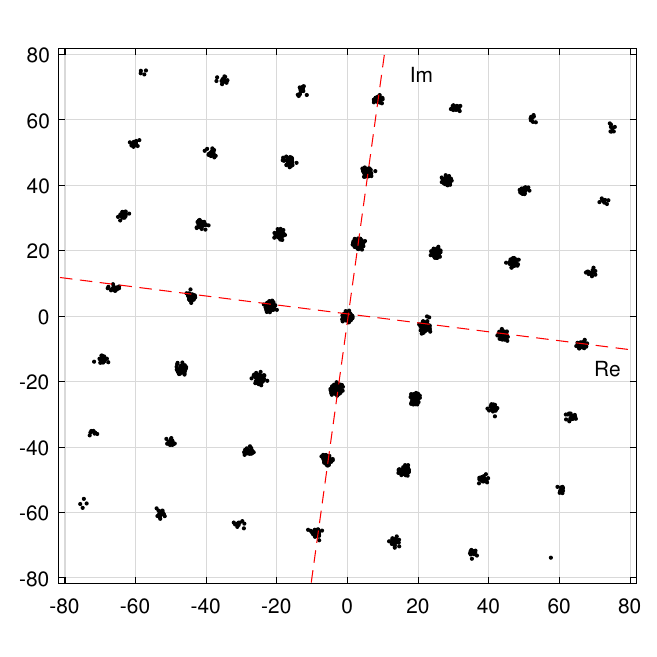}
		\label{fig:beamforming}
	}\hfil
	\subfloat[SLP]{
		\includegraphics[width=0.25\textwidth]{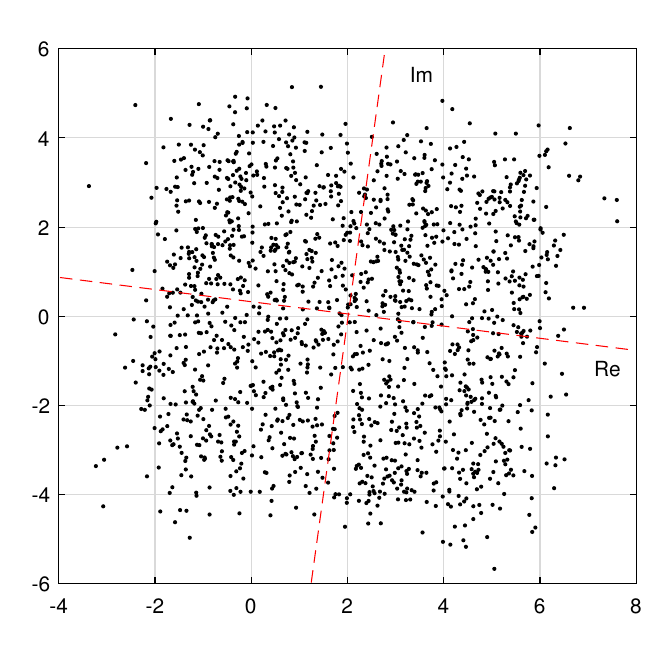}
		\label{fig:slp16}
	}\hfil
	\subfloat[ISCC]{
		\includegraphics[width=0.25\textwidth]{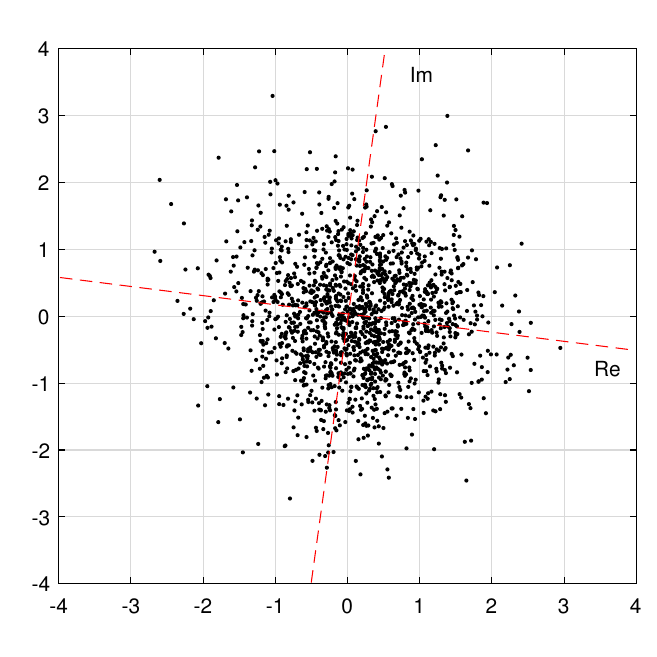}
		\label{fig:secslp16}
	}
	\caption{The constellation of received signals at targets under 16QAM.}
	\label{fig:constellation_QAM}
\end{figure*}
Similar to the phenomena observed with PSK modulation, as shown in Fig.~\ref{fig:constellation_QAM}, analogous behaviors are observed under 16QAM modulation. The beamforming approach results in overlapping 16QAM constellations at the eavesdropper's receiver, while the SLP maintains a distinguishable 16QAM structure. In contrast, the proposed ISCC method effectively obscures the constellation, producing a uniform, noise-like distribution at the malicious target's receiver, thereby enhancing covertness.

\subsection{Imperfect CSI Case}

Our preceding developments can also be extended to the case of imperfect CSI. Fig.~\ref{fig:beamet} illustrates the beampatterns under different angular deviation bounds $\varepsilon_T$, corresponding to $1^\circ$, $3^\circ$, $5^\circ$, and $6^\circ$. The targets are located at $\theta_{t0}$ = $-15^{\circ}$ and $\theta_{t1}$ = $15^{\circ}$. And the CUs are located at $\theta_{k0}$ = $-40^{\circ}$ and $\theta_{k1}$ = $40^{\circ}$. When the uncertainty is small (e.g., $\varepsilon_T=1^\circ$), the mainlobe is sharp and the sidelobe levels are low, indicating accurate beam focusing. As $\varepsilon_T$ increases, the beampattern gradually broadens and the mainlobe gain decreases slightly, reflecting the robustness introduced to accommodate larger angular errors. 
These results confirm the effectiveness of the robust design in maintaining stable beampatterns under different levels of angle uncertainty.

The robust performance under both PSK and QAM signaling is illustrated in Figs.~\ref{fig:robustpsk} and \ref{fig:robustqam}. In the case of QPSK, Fig.~\ref{fig:robustpsk}(a) compares the SER performance under channel uncertainty with and without robust design. When robustness is not considered, the SER clearly degrades as the uncertainty bound increases. In contrast, the robust ISCC design maintains significantly lower SER, showing strong resilience against channel mismatch. While in Fig.~\ref{fig:robustpsk}(b) the SCNR at malicious targets decreases slightly as $\varepsilon_U$ grows, revealing a fundamental trade-off between communication reliability and sensing accuracy. Similar trends are observed under 16-QAM signaling in Fig.~\ref{fig:robustqam}.
\addtocounter{figure}{1}
\begin{figure}[!h]
	\centering
	\subfloat[SER performence]{
		\includegraphics[width=0.7\columnwidth,trim=10 2 20 22,clip]{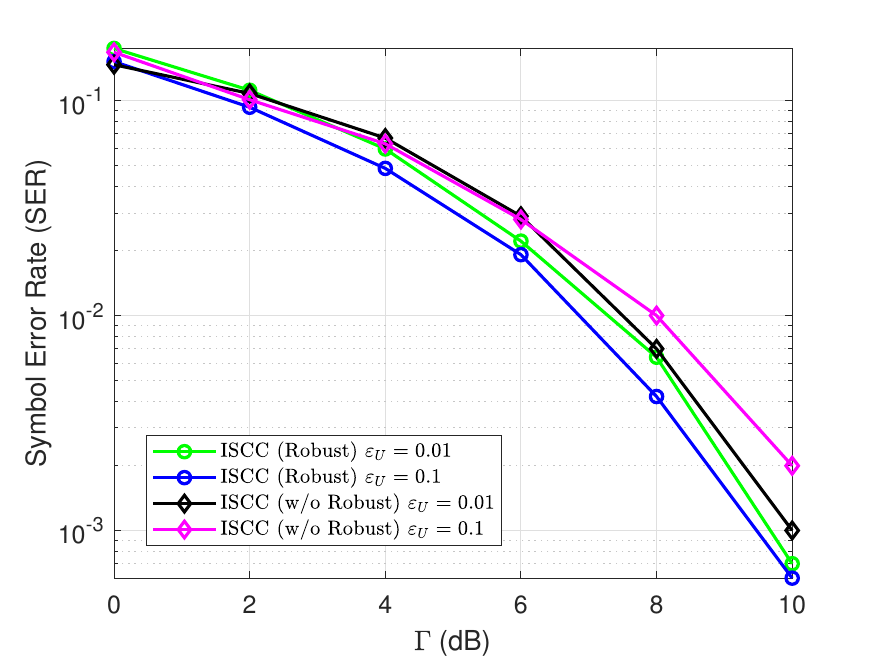}
	}\hfill
	\subfloat[SCNR performence]{
		\includegraphics[width=0.7\columnwidth,trim=10 2 20 19,clip]{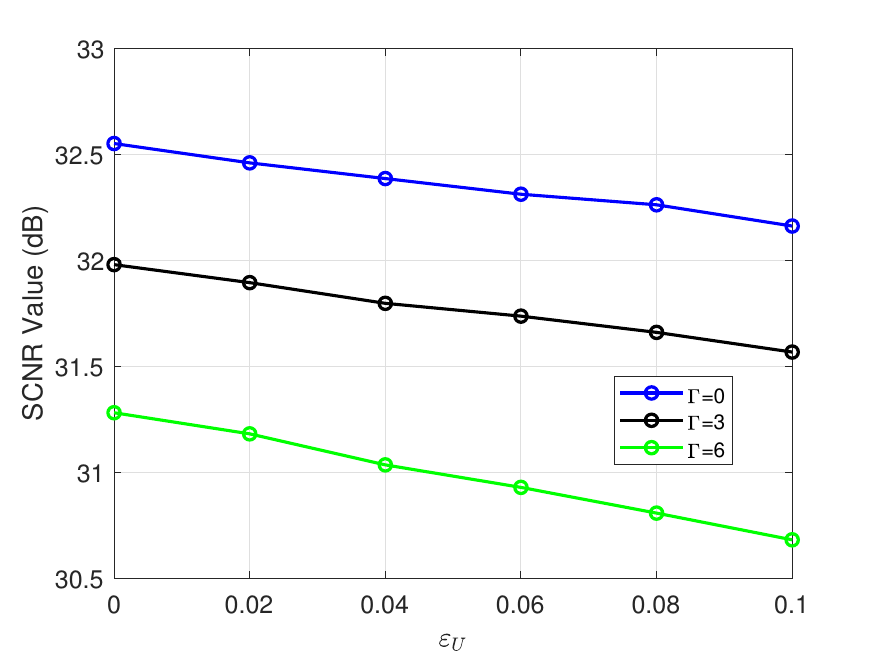}
	}
	\caption{Robust performance under different ${\varepsilon}_{U}$ (QPSK).}
	\label{fig:robustpsk}
\end{figure}

\begin{figure}[!h]
	\centering
	\subfloat[SER performence]{
		\includegraphics[width=0.7\columnwidth,trim=10 2 20 22,clip]{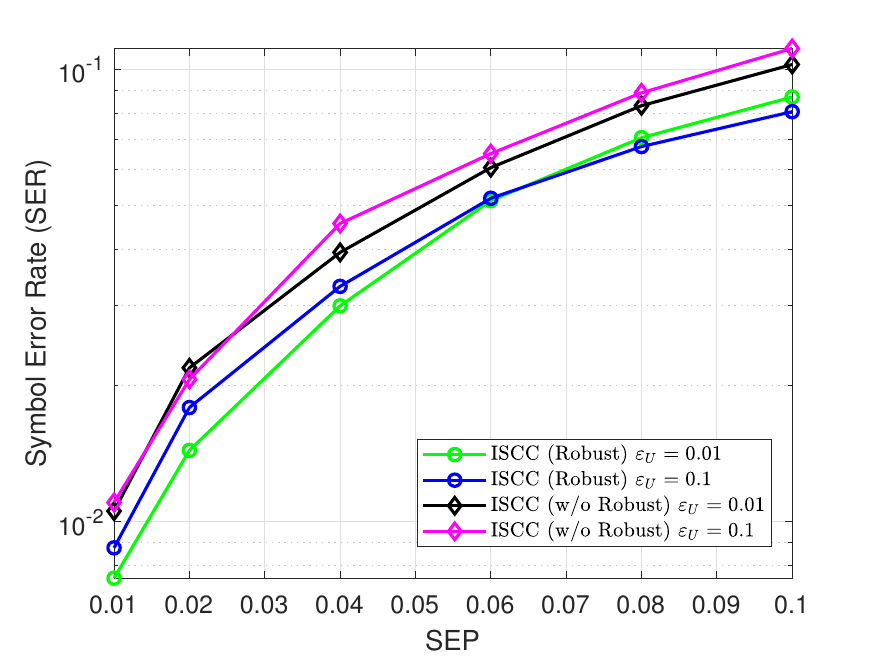}
	}\hfill
	\subfloat[SCNR performence]{
		\includegraphics[width=0.7\columnwidth,trim=10 2 20 19,clip]{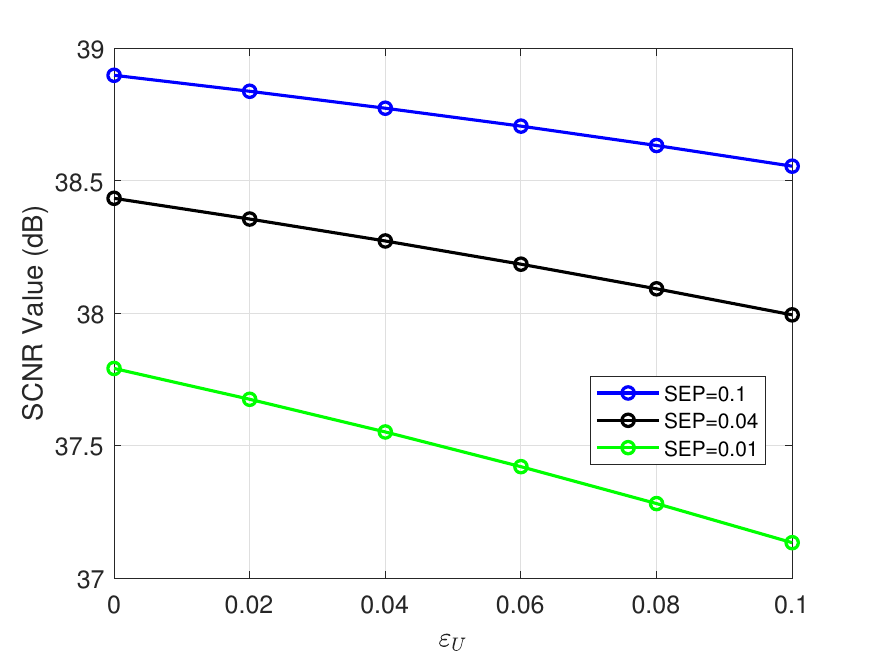}
	}
	\caption{Robust performance under different ${\varepsilon}_{U}$ (16QAM).}
	\label{fig:robustqam}
\end{figure}

\section{Conclusions} \label{sec:conclusions}
In this paper, we proposed a novel symbol-level waveform optimization framework for integrated sensing and covert communication (ISCC), addressing the trade-offs between sensing performance, communication reliability, information security, and covertness. To tackle the resulting non-convex problem, we developed a low-complexity proximal distance algorithm. Numerical results demonstrate that the proposed approach outperforms traditional beamforming and SLP designs without noise shaping, achieving improved communication performance and enhanced security and covertness. Furthermore, the incorporation of noise shaping effectively impairs eavesdropping capabilities and maintains a consistently high symbol error rate at the malicious targets. In addition, we extended the framework to robust ISCC optimization under imperfect CSI, ensuring worst-case guarantees against user and target uncertainties.

For future work, it would be of interest to investigate RIS-aided ISCC, where the joint design of transmit waveforms and RIS reflection phases could further enhance sensing and communication performance. Another promising direction lies in distributed or cooperative ISAC networks, where multiple transmitters collaboratively optimize their waveforms to balance sensing reliability and communication covertness.

 \appendix
 
\subsection{Projection onto \eqref{eq:mm_main_epi_c1} and \eqref{eq:mm_main_epi_c2} }\label{appendix:project_1}
 
 Let ${\cal L}(\bm z,  \lambda) =  \| \bm z - { \bar{\bm x}_{\ell}}\|^2  - \lambda (\Re\{ \tilde{\bm h}_k^H \bm z\} - \mu_k)$ be the Lagrangian function and $\lambda\in \mathbb{R}_+$ be the Lagrangian multiplier. Then, the optimal $\bm z^\star$ for the above projection is given by
 \begin{equation}
 	\bm z^\star = { \bar{\bm x}_{\ell }} + \frac{\lambda }{2}  \tilde{\bm h}_k, 
 \end{equation}  
 where the optimal $\lambda$ should satisfy $\lambda (\Re\{ \tilde{\bm h}_k^H \bm z^\star\} - \mu_k) = 0$. There are two cases: 
 \begin{enumerate}[label=(\alph*)]
 	\item $\lambda =0$, which means that  $\bm z^\star = { \bar{\bm x}_\ell}$ is optimal  and satisfies $\Re\{ \tilde{\bm h}_k^H \bm z^\star\} \geq \mu_k$. Simply speaking, the point $ { \bar{\bm x}_\ell}$ is already in the set ${\cal C}_{\ell,k}$ and the projection is ${ \bar{\bm x}_\ell}$ itself.
 	\item $\lambda>0$, in such a case we need to find $\lambda$ such that  $\Re\{ \tilde{\bm h}_k^H \bm z^\star\} - \mu_k = 0$. It is easy to see that the $\lambda$ is given by
 	\begin{equation}
 		\lambda = 2(\mu_k  - \Re\{ \tilde{\bm h}_k^H \bar{\bm x}_\ell\} ) /\| \tilde{\bm h}_k\|^2. 
 	\end{equation} 
 \end{enumerate}
 To summarize, the optimal $\bm z^\star$ for problem~\eqref{eq:proj_1_1} is given by
 \begin{equation} 
 	\bm z^\star = { \bar{\bm x}_\ell} +  \frac{\left(\mu_k  - \Re\{ \tilde{\bm h}_k^H \bar{\bm x}_\ell\} \right)^+}{ \| \tilde{\bm h}_k\|^2}  \tilde{\bm h}_k.
 \end{equation} 

\subsection{Detailed procedure for identifying $\tau^\star$} \label{appendix_search_tau}
 Notice that the constraints in~\eqref{eq:proj_un_d_eq_seq_b} implicitly impose $\tau\geq \frac{a_\ell+c_\ell}{2},~\forall~\ell\in {\cal L}$. Hence, the  feasible domain of $\tau$ is $\tau\geq \tau_{\min}$ where $\tau_{\min} = \max\{ 0, \frac{a_\ell+c_\ell}{2}\forall\ell\}$. We define the  breakpoint set \(\tilde{\mathcal{D}}\) over this domain as
 $
 	\tilde{\mathcal{D}} \triangleq \{ \tau_{\min}\} \cup \{\tau_{\ell,1}~|~\tau_{\ell,1}\geq \tau_{\min}, \forall\ell\}   \cup \{\tau_{\ell,2}~|~\tau_{\ell,2}\geq  \tau_{\min},~\forall\ell\} \cup \{+\infty\}.    
$
 Sorting the elements of $\tilde{\mathcal{D}}$ in ascending order yields the partition $\mathcal{D} = \{\omega_1, \ldots, \omega_{|\mathcal{D}|}\}$, which divides the feasible domain of $\tau$ into $|\mathcal{D}|-1$ intervals $[\omega_p, ~\omega_{p+1}]$. Within each interval, the objective in~\eqref{eq:proj_un_d_eq_seq_a} becomes piecewise quadratic and can be minimized efficiently. Specifically, for each interval $[\omega_p, ~\omega_{p+1}]$, we solve the following subproblem:
 \begin{equation}
 	\begin{aligned}
 		\min_{\tau} & ~ \sum_{\ell\in\mathcal{T}_p}[u_\ell(\tau) - \bar{\upsilon}_\ell]^2 + \sum_{\ell\in\mathcal{L}_p}[l_\ell(\tau) - \bar{\upsilon}_\ell]^2 + (\tau-\bar{\tau})^2 \\
 		\rm{s.t.} & ~ \omega_p \leq \tau \leq \omega_{p+1},
 	\end{aligned}
 \end{equation}
 where ${\cal T}_p = \{ \ell~|~u_\ell (\omega_{p+1}) \leq \bar{\upsilon}_\ell,~ s_\ell \in \mathbb{R}_+\} \cup  \{ \ell~|~u_\ell (\omega_{p}) \leq \bar{\upsilon}_\ell, s_\ell \in \mathbb{R}_-\} $ and $ {\cal L}_p = \{ l_\ell (\omega_{p+1}) \leq \bar{\upsilon}_\ell,~s_\ell \in \mathbb{R}_+\}  \cup  \{ l_\ell (\omega_{p}) \leq \bar{\upsilon}_\ell,~s_\ell \in \mathbb{R}_-\}.$ This problem admits a closed-form solution:
 \begin{equation}
 	\tau^p = \max\{ \omega_p , ~\min\{ \omega_{p+1}, \tilde{\tau}^p\}\},
 \end{equation}
 where 
 \begin{equation}
 	\tilde{\tau}^p = \frac{\sum\limits_{\ell\in\mathcal{T}_p}(s_\ell+1)(c_\ell+\bar{\upsilon}_\ell) - \sum\limits_{\ell\in\mathcal{L}_p}(s_\ell-1)(a_\ell-\bar{\upsilon}_\ell) + \bar{\tau}}{1 + \sum\limits_{\ell\in\mathcal{T}_p}(s_\ell+1)^2 + \sum\limits_{\ell\in\mathcal{L}_p}(s_\ell-1)^2}.
 \end{equation}
 The corresponding objective value is
 \begin{equation} \label{eq:optimal_d}
 	\begin{aligned}
 		f^p = \sum_{\ell\in\mathcal{T}_p}[u_\ell(\tau^p) - \bar{\upsilon}_\ell]^2 + \sum_{\ell\in\mathcal{L}_p}[l_\ell(\tau^p) - \bar{\upsilon}_\ell]^2 + (\tau^p-\bar{\tau})^2.
 	\end{aligned}
 \end{equation}
 Finally, the global optimal $\tau^\star$ of problem~\eqref{eq:proj_un_d_eq_seq} is obtained by selecting the candidate \(\tau^p\) that achieves the minimum \(f^p\) across all intervals.

\subsection{Proof of Proposition~\ref{prop:robust_sep_eqv}}\label{sec:proof_prop1}
 
 Consider~\eqref{eq:robust_psk_b}–\eqref{eq:robust_psk_c}.  
 For $\bm p=\bm x_\ell s$ with $|s|=1$, we have
 \begin{equation}
 \begin{aligned}
 	& \min_{\|\Delta\bm h_k\|_2\le\varepsilon_{U,k}} \Re\{(\hat{\bm h}_k+\Delta\bm h_k)^H \bm p\}\\
  = &  \Re\{\hat{\bm h}_k^H \bm p\} + \min_{\|\Delta\bm h_k\|_2\le\varepsilon_{U,k}} \Delta\bm h_k^H \bm p \\
  	= & \Re\{\hat{\bm h}_k^H \bm p\} - \varepsilon_{U,k}\|\bm p\|_2 \\
  = 	&\Re\{\hat{\bm h}_k^H \bm x_\ell s\}-\varepsilon_{U,k}\|\bm x_\ell\|_2,
 \end{aligned}     
 \end{equation}
 where the last step follows from the Cauchy--Schwarz inequality.  
 Thus both robust constraints are equivalent to
\begin{equation}
    \Re\{\hat{\bm h}_k^H \bm x_\ell s\}-\varepsilon_{U,k}\|\bm x_\ell\|_2 \ge \mu_k,
 \qquad s\in\{\tilde{s}_{k,\ell},\bar{s}_{k,\ell}\}.
\end{equation}
 Introducing a slack $r_\ell\ge 0$ gives the SOC system in~\eqref{eq:robust_sep_soc}, completing the proof.

\bibliographystyle{IEEEtran}

\vspace{-30pt}
\begin{IEEEbiography}[{\resizebox{.9in}{!}{\includegraphics{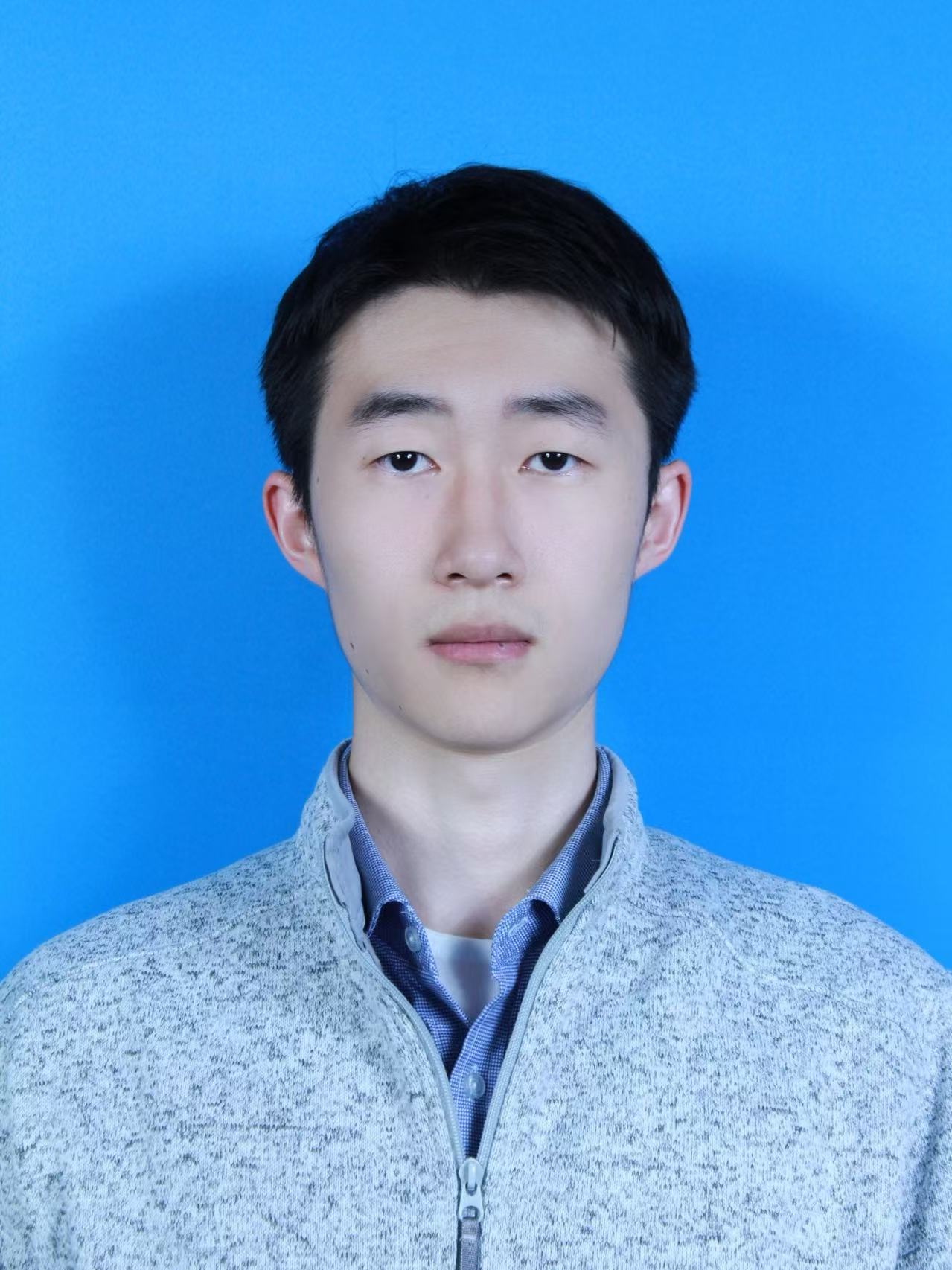}}}]
	{Yufei Wang} received the B.Eng. and M.Eng. degrees from the University of Electronic Science and Technology of China, Chengdu, in 2022 and 2025, respectively.  His research interests mainly focus on wireless communications and integrated sensing and communication. He is currently with Huawei Technologies Co., Ltd., with a focus on embedded development for communication chips.
\end{IEEEbiography}
\vspace{-30pt}
\begin{IEEEbiography}[{\resizebox{.94in}{!}{\includegraphics{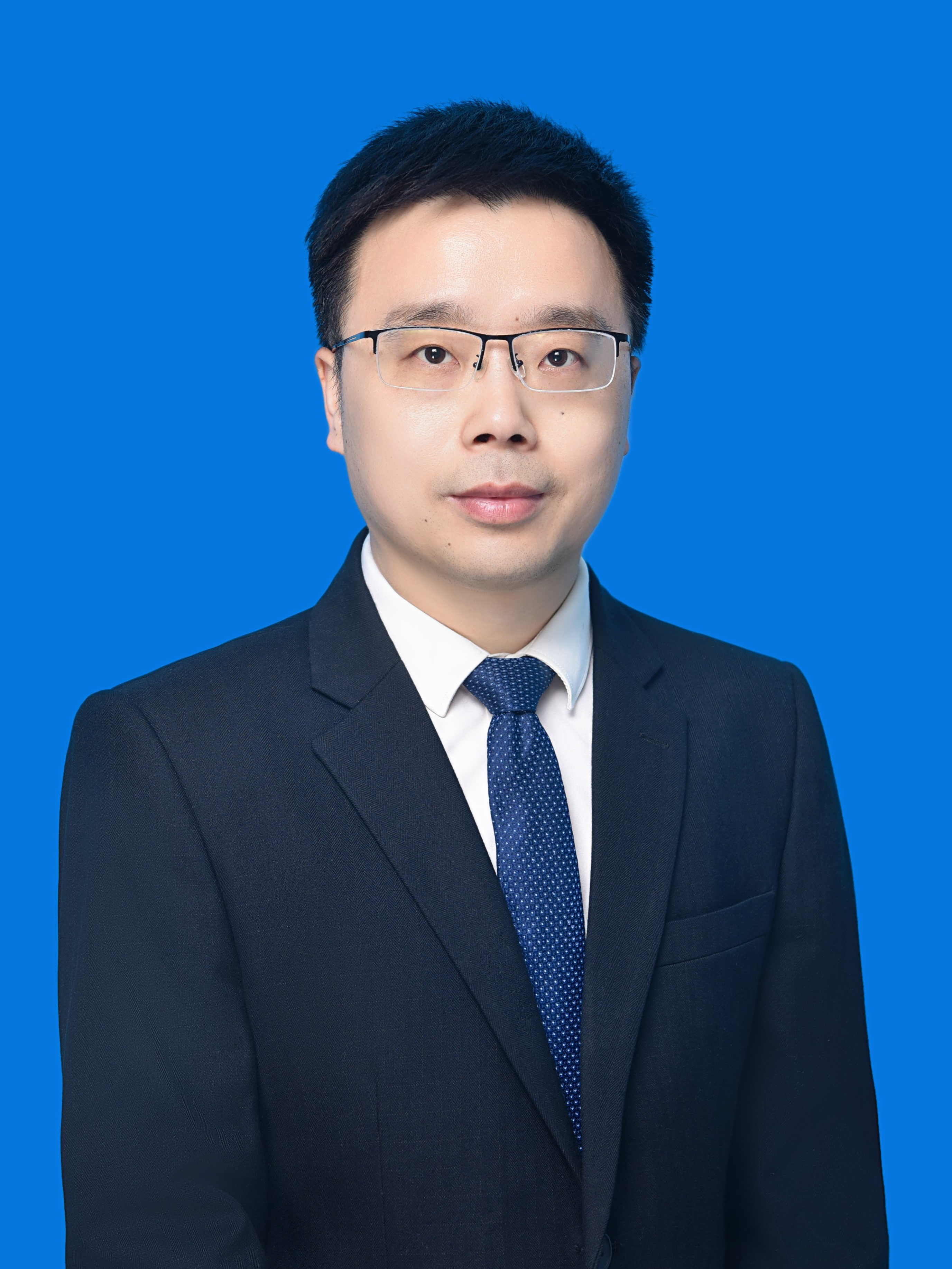}}}]
	{\bf Qiang Li} received the B.Eng. and M.Phil. degrees in Communication and Information Engineering from University of Electronic Science and Technology of China (UESTC), Chengdu, China, and the Ph.D. degree in Electronic Engineering from the Chinese University of Hong Kong (CUHK), Hong Kong, in 2005, 2008, and 2012, respectively. He was a  Visiting Scholar with the University of Minnesota and Research Associate with the Department of Electronic Engineering and the Department of Systems Engineering and Engineering Management, CUHK. Since November 2013, he has been with the School of Information and Communication Engineering, UESTC, where he is currently a Professor. His recent research interests focus on machine learning and intelligent signal processing	in wireless communications.	He received a Best Paper Award of IEEE PIMRC 2016, and the Best Paper Award of the IEEE Signal Processing Letters 2016.
\end{IEEEbiography}
\vspace{-30pt}
\begin{IEEEbiography}[{\resizebox{.94in}{!}{\includegraphics{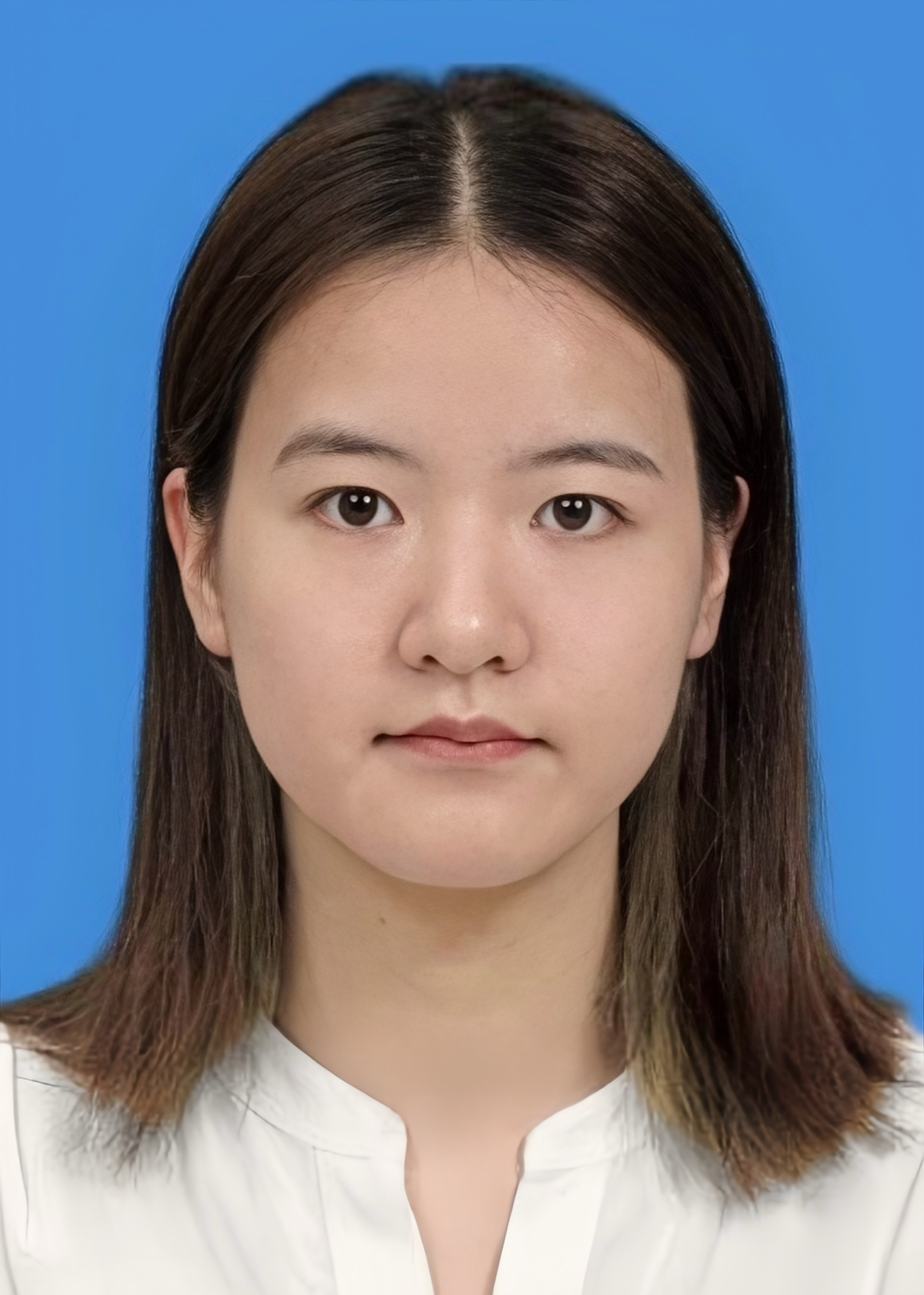}}}]
	{\bf Hongli Liu} received the B.Eng. degree from the Southwest University, Chongqing, China, in 2022. She is currently working toward the Ph.D. degree at the School of Information and Communication Engineering, University of Electronic Science and Technology of China, Chengdu, China. Her research interests are mainly on wireless communications, massive MIMO systems, and integrated sensing and communication.
\end{IEEEbiography}
\vspace{-30pt}
\begin{IEEEbiography}[{\resizebox{.94in}{!}{\includegraphics{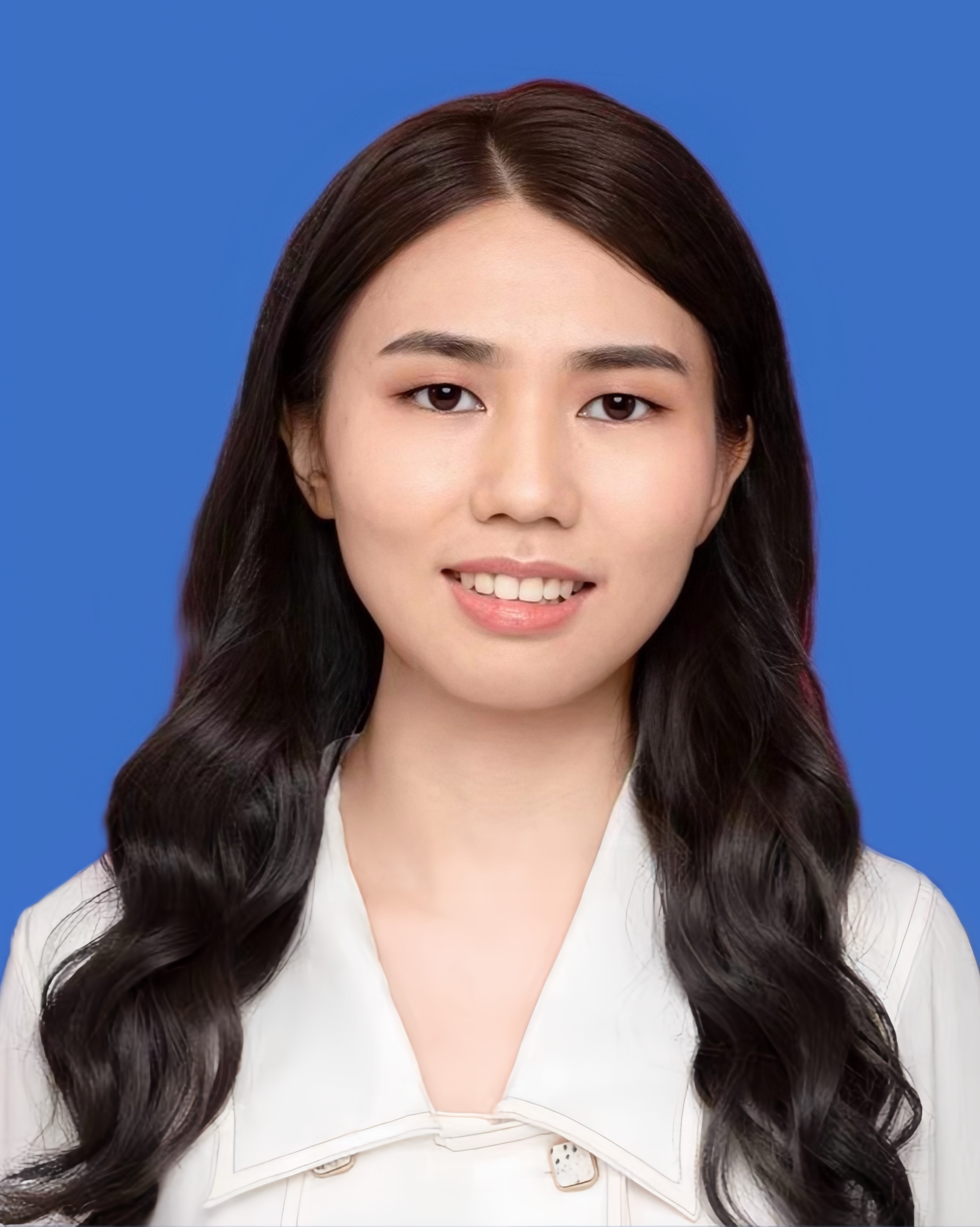}}}]
	{Ying Zhang} received the B.Eng. degree from Chongqing University of Posts and Telecommunications, Chongqing, China, in 2014, and the M.S. degree in electronic and communication engineering from the University of Electronic Science and Technology of China (UESTC), Chengdu, China, in 2017. She is currently pursuing the Ph.D. degree at the School of Information and Communication Engineering, UESTC. Her research interests include wireless communications and signal processing.
\end{IEEEbiography}
\vspace{-30pt}
\begin{IEEEbiography}[{\resizebox{.94in}{!}{\includegraphics{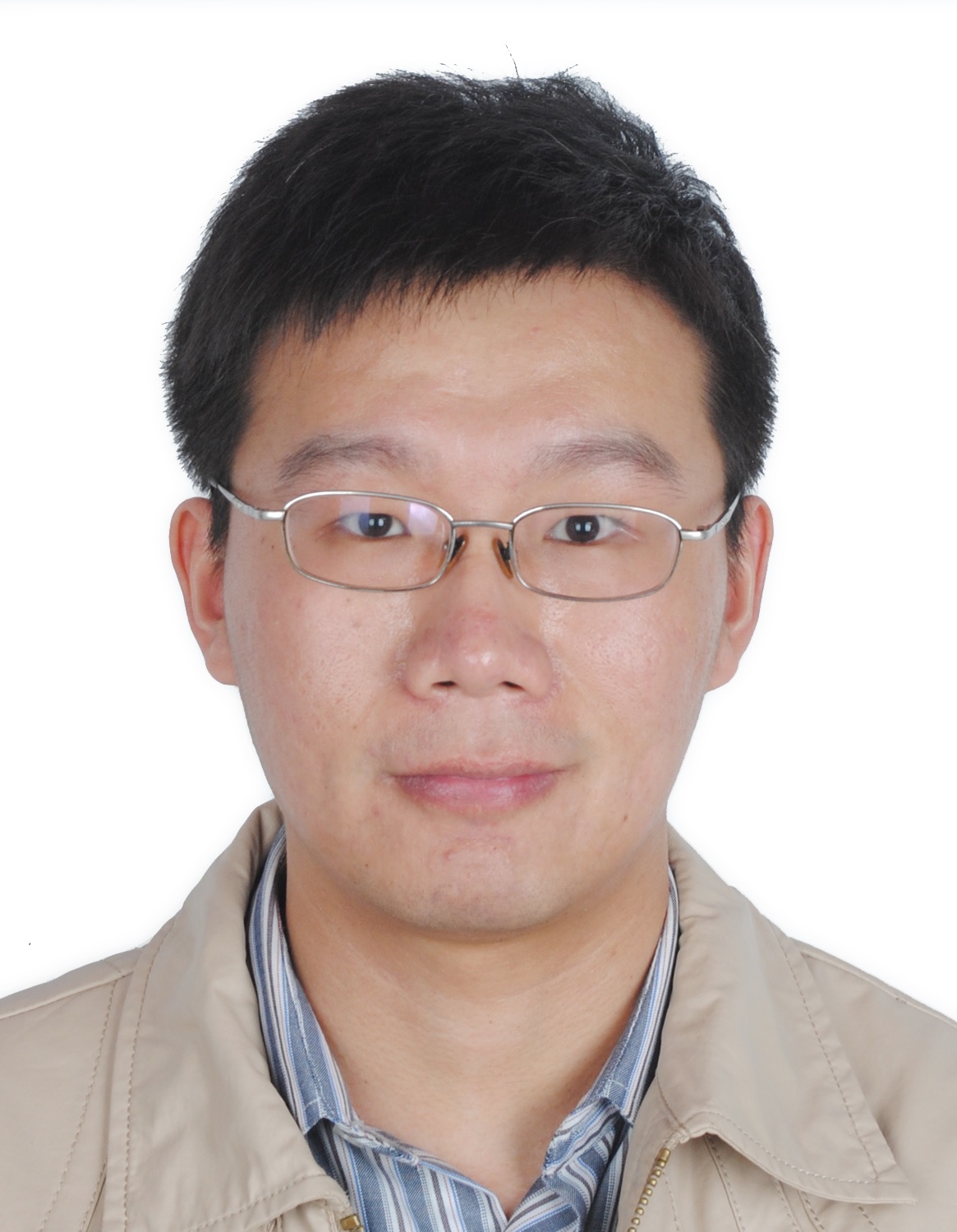}}}]
	{\bf Jingran Lin}  received the B.S. degree in Computer Communication from University of Electronic Science and Technology of China (UESTC), Chengdu, China, in 2001, and the M.S. and Ph.D. degrees in Signal and Information Processing from UESTC in 2005 and 2007, respectively.	After his graduation in June 2007, he joined the School of Information and Communication Engineering, UESTC, where he is currently a Full Professor. From January 2012 to January 2013, he was a Visiting Scholar with the University of Minnesota (Twin Cities), Minneapolis, MN, USA. His research interests include the algorithm design and analysis for the intelligent signal processing problems arising from modern communication systems.	
\end{IEEEbiography}

\end{document}